\documentclass[10pt,preprint2]{emulateapj}
\usepackage{natbib}

\shorttitle{VINE - a new hybrid SPH / $N$-body code}
\shortauthors{Wetzstein et al.}

\begin{document}

\newcommand{\vineI}{Vine1}
\newcommand{\vineII}{Vine2}

\newcommand{\cpc}{Comp. Phys. Comm.}
\newcommand{\cpr}{Comp. Phys. Rep.}
\newcommand{\siamcomp}{SIAM J. Scient. Comp.}
\newcommand{\jcomp}{J. Comp. Phys.}
\newcommand{\ti}{\textit}
\newcommand{\m}{\mathbf}
\newcommand{\f}{\frac}
\newcommand{\beq}{\begin{equation}}
\newcommand{\eeq}{\end{equation}}
\newcommand{\beqa}{\begin{eqnarray}}
\newcommand{\eeqa}{\end{eqnarray}}
\newcommand{\erfc}{\mbox{erfc}}
\newcommand{\erf}{\mbox{erf}}

\title{VINE -- A numerical code for simulating astrophysical systems
using particles I: Description of the physics and the numerical methods}

\author{M. Wetzstein\altaffilmark{1,2}, Andrew F. Nelson\altaffilmark{3,4},
  T. Naab\altaffilmark{2,5} and A. Burkert\altaffilmark{2}}
\email{mwetz@usm.lmu.de}

\altaffiltext{1}{Department of Astrophysical Sciences, Princeton University,
                     Princeton, NJ 08544, USA}
\altaffiltext{2}{Universit\"ats-Sternwarte, Scheinerstr. 1,
                     81679 M\"unchen, Germany}
\altaffiltext{3}{Los Alamos National Laboratory, HPC-5 MS B272,
                  Los Alamos NM, 87545, USA}
\altaffiltext{4}{UKAFF Fellow}
\altaffiltext{5}{Institute of Astronomy, Maddingley Road, Cambridge,
                      United Kingdom}

\begin{abstract}
We present a numerical code for simulating the evolution of
astrophysical systems using particles to represent the
underlying fluid flow. The code is written in Fortran 95 and is designed to
be versatile, flexible and extensible, with modular options
that can be selected either at the time the code is compiled or at run
time through a text input file. We include a number of general purpose
modules describing a variety of physical processes commonly required
in the astrophysical community and we expect that the effort required
to integrate additional or alternate modules into the code will small.
In its simplest form the code can evolve the dynamical
trajectories of a set of particles in two or three dimensions using a
module which implements either a Leapfrog or Runge-Kutta-Fehlberg
integrator, selected by the user at compile time. The user may choose
to allow the integrator to evolve the system using individual
timesteps for each particle or with a single, global time step for
all. Particles may interact gravitationally as $N$-body particles, and
all or any subset may also interact hydrodynamically, using the
Smoothed Particle Hydrodynamic (SPH) method by selecting the SPH
module. A third particle species can be included with a module
to model massive point particles which may accrete nearby SPH or
$N$-body particles. Such particles may be used to model, e.g., stars
in a molecular cloud. Free boundary conditions are implemented by
default, and a module may be selected to include periodic boundary
conditions. We use a binary `Press' tree to organize particles for
rapid access in gravity and SPH calculations. Modules implementing an
interface with special purpose `GRAPE' hardware may also be selected
to accelerate the gravity calculations. If available, forces obtained
from the GRAPE coprocessors may be transparently substituted for those
obtained from the tree, or both tree and GRAPE may be used as a
combination GRAPE/tree code. The code may be run without modification
on single processors or in parallel using OpenMP compiler directives
on large scale, shared memory parallel machines. We present simulations of
several test problems, including a merger simulation of two elliptical
galaxies with $800000$ particles. In comparison to the
Gadget-2 code of \citet{springel_g2}, the gravitational force
calculation, which is the most costly part of any simulation including
self-gravity, is $\sim 4.6 - 4.9$ times faster with VINE when
tested on different snapshots of the elliptical galaxy merger simulation
when run on an Itanium~2 processor in an SGI Altix. A full
simulation of the same setup with $8$ processors is a factor
of $2.91$ faster with VINE. The code is
available to the public under the terms of the Gnu General Public License. 
\end{abstract}



\section{Introduction}

In modern astrophysics, the numerical simulation of systems whose
complexity is beyond the capabilities of analytical models has become
a widely used tool. On nearly all length scales, ranging from problems
on cosmological distances, to galaxy formation and evolution to star
and planet formation, numerical simulations have contributed much to
our current understanding of the physical processes which lead to the
universe we observe.

The numerical simulation of a self-gravitating and/or gas dynamical
system is a basis common to all those problems, no matter what length
scale they belong to. The simulation techniques for such systems can
be divided into two different approaches: grid based methods divide
space into finite sized cells and compute the physical quantities such
as temperature, pressure, etc., inside those cells \citep[see e.g.][ and
references therein]{stone1992, ryu1993, oshea2004}. Particle based
methods represent a system by a set of particles to which physical
quantities such as mass, position and velocity are assigned or
computed \citep[see e.g.][ and references therein]{hern_katz89,
dave97, springel2001, wadsley2004, springel_g2}. Which approach is
best for modeling a particular system depends both on the problem to
be modeled and the biases of the researcher doing the modeling.
Without going into the details of relative merits and shortcomings of
either approach, we point out that for some problems, a grid based
approach may be nearly unfeasible because of the existence of
irregular boundaries. Large voids can also be problematic for a grid
based simulation because they require a large number of empty or
nearly empty and uninteresting zones be included, at a high
computational expense. Collisionless systems are typically only
modelled with particle representations. Although these can be modelled  
using a grid method (i.e. 'Particle In Cell', or PIC,
methods) to calculate a long range force such as gravity, these
systems are often modelled using tree-based or direct summation forces. 
A particle based simulation
naturally concentrates the computational work in the most interesting
areas of the flow, in most cases a very valuable feature, but the
absence of particles in voids can be problematic if the simulation
requires such low density regions to be resolved at high accuracy
\citep[see e.g.][ for a recent comparison of grid and particle based
codes for cosmological simulations]{oshea2005}. It may also suffer
from a relatively poorer reproduction of the fluid behavior at shocks.

For a system evolving only under the influence of gravity, a particle
based approach leads to the classical $N$-body problem. A set of $N_p$
particles evolve according to the force on each particle exerted by
all the others. The efficient computation of these forces is a
longstanding numerical problem. If gas dynamics is also required, the
Smoothed Particle Hydrodynamic (SPH,
\citealp{lucy77,ging_mon77,benz90, monaghan92}) method has achieved
great success at incorporating such processes into the framework of a
particle method. In SPH, the gas is also represented using particles
(which makes the method so useful in combination with $N$-body
methods), which are assumed to sample the local hydrodynamic
quantities of the underlying flow. In addition to a position and
velocity, these particles also possess an internal energy intrinsic to
each and a volume (or surface) density that is reconstructed for each
particle, based on the positions of nearby `neighbor' particles. Thus,
these gas particles feel not only the gravitational forces that all
particles in the simulation do, but also pressure forces and other
gas dynamical effects.

Modeling dynamical and hydrodynamical systems using particles relies
on the sufficiently accurate computation of both the gravitational and
hydrodynamical forces of the particles on each other, then advancing
them forward in time according to those forces. Thus a time
integration of the particles' equations of motion and additional
equations for hydrodynamic quantities such as internal energy, is the
problem to be solved. Constraints on the time integration are that we
would like it to faithfully reproduce the evolution of the real system
that the simulation is supposed to model, and that it do so
efficiently, so that results may be obtained quickly and insight into
the physical world gained at a minimum of cost.

In this paper and a companion \citep*[][ hereafter \vineII]{vineII}, we
describe a numerical code for efficiently simulating the evolution of
astrophysical systems using $N$-body particles, with the optional
additions of including gas dynamical effects using the SPH method,
self gravity and additional massive `star' particles which may accrete
the other species of particles. We call this code VINE. The
present paper describes the physics we have implemented, the high
level code design and the results of simulations using the code on a
number of test problems as well as a comparison to the
Gadget-2 code of \citet{springel_g2}. \vineII\  describes the low level 
design and optimization of the most computationally expensive parts of
the code, the methods used to parallelize it and the performance of
each part in serial and in parallel. VINE has been succesfully used
for a large series of simulations of galaxy interactions
\citep{naab2003, jesseit2005, burkert2005, dasyra2006, dasyra2006b,
naab2006, naab2006b, naab2006c, bell2006, thomas2007, burkert2007,
naab2007, jesseit2007, wetzstein2007}, turbulence studies 
\citep{kitsionas2008}, as well as planet formation
\citep[e.g.][]{nelson06}. In addition, VINE has been extended with
an implementation of ionizing radiation in SPH \citet{gritsch2008, 
gritsch2009}.

In section \ref{sec:integ} we describe the implementation of two
second order integration methods included with the code, and a
discussion of the criteria used to determine the time steps used to
evolve the particles forward in time. We describe in sections
\ref{sec:sph} and \ref{sec:gravity}, respectively, the form of the
equations used to implement the SPH method and the different options
for the calculation of gravitational forces. In section \ref{sec:PM},
we describe the implementation of `star' particles which can accrete
the $N$-body and SPH particle species. The boundary conditions
available in the code are discussed in section \ref{sec:bounds}. In
section \ref{sec_tests} we demonstrate some of the capabilities of the
code on several test simulations, including both SPH as well as a pure
$N$-body problems. In section \ref{sec:perf} the performance of VINE
is compared to that of the Gadget-2 code of \citet{springel_g2}.
Finally, in section \ref{sec_summary}, we summarize the features of
our code and give web sites where the source code may be obtained
electronically. 


\section{Time integration}\label{sec:integ}

In order to simulate the evolution of a physical system using a set of
$N_p$ particles, we require first a set of equations by which the system
evolves and second a method for integrating those equations forward in
time. In the case of particle systems involving only gravitational
interactions (`$N$-body' simulations), the equations form a set of
coupled first order differential equations governing the motion of
those particles in response to each other and (if present) to outside
influences. This set consists of the equations describing the motion
of each particle, 
\begin{equation}\label{eq:pos-deriv}
{d{\mathbf x}_i\over{dt}} = {\mathbf v}_i 
\end{equation}
and the equations for momentum conservation
\begin{equation}\label{eq:vel-deriv}
{d{\mathbf v}_i\over{dt}} = -{{{\mathbf \nabla} \Phi}\over{m}_i}.
\end{equation}
The solution of these equations is difficult because each of the
$4N_p$ or $6N_p$ equations (in 2 or 3 dimensions) is coupled through
the gravitational potential, $\Phi$. Other coupling terms must be added
in cases where other physical phenomena such as hydrodynamics (section
\ref{sec:sph}) are active, and may require additional equations be
solved. For the case of hydrodynamic systems, we must solve not only a
momentum equation appropriately generalized from above, but also mass
and energy conservation equations.

A variety of methods for integrating differential equations are
available \citep[see~e.g. the textbooks of ][]{hockney81,Fletcher},
with varying degrees of utility for any given problem. Determining
which method is most efficient is highly dependent on the
characteristics of the problem itself. Most astrophysical systems, for
example, develop highly non-linear flow patterns with the practical
consequence that high order integration methods (i.e. those with a
mathematical truncation error proportional to a high power of the
integration step size) are not generally useful. The nonlinearities
mean that time steps must be restricted to very small sizes in order
to resolve the flow, while the high order integration requires many
derivative calculations per timestep, yielding a very high
computational cost to evolve a system for a given amount of time. Even
among integrators of identical `order', characteristics of the errors
that develop can vary, with one being inappropriate for use on a
problem another might be ideally suited to solve. 

In order to allow the user enough flexibility to determine what is
best for a specific problem, we have implemented both a second order
Runge-Kutta scheme \citep{fehlberg68} and a second order leapfrog
scheme \citep[see
e.g.][]{hockney69,hern_katz89,rasio_shapiro91,springel2001} in VINE.
Although very different in structure (leapfrog requires only one force
computation per time step for example, while the Runge-Kutta scheme
requires two), users may transparently select one or the other at the
time the code is compiled. Both integrators are very modular in the
sense that they use the same bookkeeping scheme for particles and
their timesteps, and identical calls to update routines. If a user
finds that still a different choice of integrator is required, we
expect that it would be straightforward to add it as an alternative as
well.

\subsection{The Runge-Kutta-Fehlberg (RKF) Integrator}\label{sec:rkf}

Runge-Kutta schemes of a variety of forms have been developed since
the original publication of the general method more than 100 years ago
\citep{kutta1901}, but until the work of \citet{fehlberg68} they
included no formal description of the size of errors that developed
during an integration. Fehlberg realized that a Runge-Kutta scheme of a
given truncation order could be embedded in a similar scheme of one
order higher, given a suitable choice of coefficients for both. The
resulting pair of methods, used together, could be used to determine a
limit on the size of the next order truncation error for the lower
order scheme in the pair. In VINE, we have implemented the first order
scheme with second order error control (`RKF1(2)'), as described by
\citet{fehlberg68}. A brief description of the scheme is summarized
here. 

For any quantity, $q$, to be integrated in time, the quantity $q^{n+1}$
at the new time $t^{n+1}$ is computed from its value $q^n$ at the
previous time $t^n$ utilizing the discretization:
\beq
\label{eq:rkf1}
q^{n+1} = q^n + (c_0 k_0 + c_1 k_1)\Delta t^n
\eeq
where the $c_i$ are constant parameters and the $k_i$ are time
derivatives of $q$ evaluated at various points during the timestep:
\beq
\label{eq:rkf_k0}
k_0 = \dot q(t^n,q^n),
\eeq
\beq
\label{eq:rkf_k1}
k_1 = \dot q(t^n + \alpha_1 \Delta t^n, q^n + \beta_{10}k_0\Delta t^n)
\eeq
\beq
\label{eq:rkf_k2}
k_2 = \dot q(t^n + \alpha_2 \Delta t^n,q^n + \beta_{20}k_0 \Delta t^n +
\beta_{21}k_1 \Delta t^n).
\eeq
where $\dot{q}$ is the time derivative of $q$ and $\Delta t^n$ is the
time step from $t^n$ to $t^{n+1}$. The $k_2$ term does not appear
directly in the integration equation \ref{eq:rkf1} above, but does
appear in the error criterion defined in section \ref{sec:rkf_ts_crit}
below. The coefficients $\alpha_k$ and $\beta_{kl}$ and $c_k$ defined
by Fehlberg are reproduced in table \ref{tab:rkf-coef}. By definition
of the coefficients, the $k_2$ term is identical to the $k_0$ term for
the following timestep, reducing the number of derivative evaluations
to two per timestep.  The $c_k$ define the coefficients used by the
first order RK scheme, while the $\hat c_k$ terms define the
coefficients used in the second order scheme used only indirectly to
define the truncation error.  

\begin{table}\label{tab:rkf-coef}
\caption{Coefficients for the RKF1(2) Integrator}
\begin{tabular}{l|l|l|l|l|l}
\hline
$k$  & $\alpha_k$  & $\beta_{k0}$ & $\beta_{k1}$ & $c_k$   & $\hat c_k$  \\
\hline
0    &   0         &              &              & 1/256   & 1/512      \\
1    &  1/2        &  1/2         &              & 255/256 & 255/256    \\
2    &   1         &  1/256       &  255/256     &         & 1/512      \\
\end{tabular}
\end{table}

\subsection{The Leapfrog Integrator}
\label{sec_leap}

The leapfrog (LF) integration scheme is formally an offset integrator:
positions and velocities are offset from each other in time by half a
time step \citep[see e.g.][]{hockney81}. Alternate updates of position
and velocity advance from one half step behind to one half step ahead
of the other update in the sequence, effectively `leapfrogging' over
each other in the integration scheme, which takes its name by analogy
from the children's game. The leapfrog implementation in VINE is
similar to that of \citet{springel2001}, for which a mathematically
equivalent form is used, in which the equations for the positions and
velocities are written in a non-offset form as 
\begin{eqnarray}
\label{eq:lf-vel}
\m{v}^{n+1} & = & \m{v}^n + \m{a}^{n+1/2} \Delta t^n  \\
\label{eq:lf-pos}
\m{x}^{n+1} & = & \m{x}^n + \f{1}{2} \left( \m{v}^n + \m{v}^{n+1}\right)
                  \Delta t^n
\end{eqnarray}
where again indices $n$, $n+1/2$ and $n+1$ denote quantities at time
$t^n$, $t^{n+1/2}$ and $t^{n+1}$, respectively, and $\Delta t^n$ is
the step from $n$ to $n+1$. To recover the offset form, notice that
positions and accelerations are actually defined on half timesteps,
but that the position update is effectively split into two halves.
With a fixed increment $\Delta t$, each position update as defined in
equation \ref{eq:lf-pos} uses the velocity corresponding to two
separate velocity updates, half from timestep $n$, $\m{v}^n/2$, and
half from timestep $n+1$, $\m{v}^{n+1}/2$, so that, effectively,
updates of position are only half completed at any `full' time step
$n$. 

The velocity update requires that accelerations be calculated on half
steps, $n+1/2$. For simulations involving self gravity and
hydrodynamics, the accelerations depend on particle positions, so that
a separate, temporary update of the position to its correctly offset
temporal location is required. This update takes the form 
\beq
\label{eq:lf_poshalf}
\m{x}^{n+1/2} = \m{x}^n + \f{1}{2} \m{v}^n \Delta t^n
\eeq
as expected from the discussion above. Other quantities requiring
integration, such as internal energy, smoothing lengths or viscous
coefficients needed for hydrodynamic simulations (section
\ref{sec:sph}), are defined on integer time steps. Their derivatives
must therefore be calculated on half time steps at the same time as
the accelerations themselves are calculated. Complications arise
because for most such variables, the derivative is a function of the
variable itself or of others defined on integer time steps. Two simple
examples are of artificial viscosity or of $PdV$ work, each of which
require velocity. VINE employs a linear extrapolation of each quantity
from $n$ to $n+1/2$, as shown in equation \ref{eq:lf_poshalf} for
position, so that the integration scheme itself remains formally
second order. In summary, the algorithm can be written as
\begin{enumerate}
\item complete position update to $\m{x}^{n+1/2}$, extrapolate other
   quantities as required.
\item compute $\m{a}^{n+1/2}$ and other derivatives.
\item update velocities $\m{v}^n \rightarrow \m{v}^{n+1}$, using
     equation \ref{eq:lf-vel}. Update other relevant quantities using
     appropriate analogous update equations.
\item update positions $\m{x}^{n} \rightarrow \m{x}^{n+1}$, using
  equation \ref{eq:lf-pos}.
\end{enumerate}
After the fourth step the sequence starts anew. This variation is
called a `Drift-Kick-Drift' (DKD) leapfrog, because particles drift
forward for half of a time step, undergo a force (the `kick')
calculation, then drift forward for an additional half step under the
control of the newly determined force. An alternative
(`Kick-Drift-Kick', or KDK) can be formulated as well, with positions
and velocities instead being defined at integer and half integer
times. This variant has been shown to have somewhat better error
properties than DKD leapfrog \citep{springel_g2}, but has not yet been
incorporated into VINE.

Although slightly more cumbersome than the pure offset form, either of
the leapfrog variants above are advantageous because adjustable time
steps, such that $\Delta t_n \not= \Delta t_{n+1}$, are
straightforward to implement, as are individual time steps for
different particles (see section \ref{sec:ind_ts}). Both features will
be desirable in simulations of systems where time scales vary widely
as conditions change over time. The consequences for such adaptability
is that the exact leapfrog symmetry between position and velocity
updates is lost, but changes between one time step value and another
should be infrequent enough in practice to make overall errors
resulting from them small.

\subsection{Timestep Criteria}\label{sec_ts_crit}

In order to produce an accurate integration, time steps must be chosen
that are small enough to maintain the stability of the system against
the growth of errors. At the same time, time steps should not be much
smaller than required to maintain stability and accuracy, because that
wastes computational resources that could be more efficiently employed
in performing larger simulations. Here we describe the criteria used
in VINE to determine timesteps for the particles.

\subsubsection{Time Step Criteria common to both the Leapfrog and RKF
Integrators}

The timestep criteria described in this section apply to both the
leapfrog and the RKF integrators. The next time step $\Delta t^{n+1}$
of a particle $i$ is determined by the minimum of value derived from a
set of $N$ criteria:
\beq 
\label{eq:ts-criteria}
\Delta t^{n+1}=\min_N(\Delta t^{n+1}_{N}),
\eeq
where we have suppressed the subscript, $i$, designating each
particle. Whether or not to include a particular criterion may be
selected by the user at compile time by commenting out (or not) calls
to subroutines that calculate one or another of the $\Delta
t^{n+1}_N$, and by routines active only when certain options are
selected, such as the Runge-Kutta integrator or SPH (sections
\ref{sec:rkf_ts_crit} and \ref{sec:tscrit-sph}).

Three simple criteria are based on changes in the acceleration of a
particle:
\beq
\label{force_crit}
\Delta t^{n+1}_{a}  =  \tau_{acc} \sqrt{\f{h}{|\m{a}|}},
\eeq
its velocity:
\beq
\label{vel_crit}
\Delta t^{n+1}_{v}  =  \tau_{vel} \f{h}{|\m{v}|},
\eeq
or both in combination:
\beq
\label{vel_acc_crit}
\Delta t^{n+1}_{va}  =  \tau_{va} \f{|\m{v}|}{|\m{a}|},
\eeq
where $h$, $\m{a}$ and $\m{v}$ are the gravitational softening length,
the acceleration and velocity of the particle $i$ at the previous time
step, respectively, and the three values of $\tau$ are tuning
parameters for each criterion. Numerical experiments show that
$\tau_{acc} \approx 0.5$ gives good results. When included, we use
similar values for the other two tolerance parameters as well.

Although the combination of all three criteria is sometimes useful,
and indeed is used in e.g. \citet{nelson06} with VINE, in many cases
it is sufficient to include only the acceleration based criterion of
equation \ref{force_crit}, allowing the others to be neglected. For
example, when the velocity criteria are included they can impose very
restrictive constraints on the calculated time step. If a particle
moves at very high velocities, equation \ref{vel_crit} can require
small time steps even when the particle does not change its trajectory
and could otherwise be integrated with large time steps. Similarly,
equation \ref{vel_acc_crit} can limit the time step of a particle when
it moves very slowly but feels only small forces. 

An additional time step criterion is also present in VINE, but is
never explicitly calculated by the code either on a particle by
particle basis or for all particles in aggregate. Instead, it is set
by the user when a specific simulation's initial conditions are
specified. Namely, the user must specify the maximum time step,
$\Delta t_{max}$, permitted by the code, on which the hierarchy of
time step bins is built. This value effectively serves as an
additonal time step criterion because it limits the steps of particles
which might obtain too large time steps through the other criteria.
Similar to the parameter values used for the integration accuracy in
the time step criteria, its optimal value chosen will be problem
dependent. In VINE, it also serves an important secondary role,
because VINE outputs checkpoint dumps for the simulation only on
integer multiples of the maximum timestep. Therefore, for an analysis
which requires high time resolution one might choose a different value
than for another analysis which takes only the final state of the
system as input.

\subsubsection{Time Step Criteria for the RKF Integrator}
\label{sec:rkf_ts_crit}

In addition to the conditions above, the RKF integrator requires an
additional criterion, which limits the second order truncation error
in the discretization (see section \ref{sec:rkf}). As defined by
Fehlberg, the second order truncation error for integrating variable
$q$ through a time $\Delta t^{(i)}$ will be:
\beq
\label{eq:rkf-truncerror}
TE  = \hat c_2 (k_0 - k_2) \Delta t^{n},
\eeq
where $k_0$ and $k_2$ are defined in equations \ref{eq:rkf_k0} and
\ref{eq:rkf_k2} and $\hat c_2$ is defined in table \ref{tab:rkf-coef}.
Unfortunately, the truncation error as defined is an absolute error.
It depends on the units for a given variable as well as the size of
the system and is therefore not particularly useful without explicit
tuning for every variable, physical system and simulation. Various
relative error metrics are straightforward to develop from equation
\ref{eq:rkf-truncerror} however. For example, we may define a relative
error metric such that the truncation error is no larger than a small
fraction of the magnitude of the variable itself:
\beq
\label{eq:rkf-normerror}
RE = {{|TE|}\over{\tau_{\mathrm{RKF}} |q|}},
\eeq
where we define $\tau_{\mathrm{RKF}}$ as a tunable parameter
restricting the error, and we require $RE$ to be $\la 1$ for an
acceptable error.

We may expect the optimal timestep to be proportional to the
square root of the truncation error, since the error itself is second
order in time. Then we may determine a new time step from the old by
comparing the ratio of the new and old values to the error metric:
\begin{equation}\label{eq:rkf-crit}
\Delta t^{n+1}_{\mathrm{RKF}} = \Delta t^{n}_{\mathrm{RKF}}
                                       \sqrt{1/RE}.
\end{equation}
With this definition, the $n+1$ time step will be decreased when the
error for a given time step is large. If small, it will be increased.
We define the final value of $\Delta t^{n+1}_{\mathrm{RKF}}$ to be the
minimum over all integration variables defined in the simulation.

Although often an improvement over the direct measure of error and
time step definition, equation \ref{eq:rkf-crit} may still suffer from
several deficiencies in practice, depending on the specific
integration variable. For example, the size of the timestep calculated
for positions depends on the position itself, and particles near the
origin will necessarily receive more restrictive time steps than
those further away. An arbitrary change of coordinate system, shifting
the entire system some distance in any direction, will also change the
error metric and time step calculation. For the same reasons, velocity
errors and timesteps will suffer similar problems.

A variety of strategies to sidestep undesirable properties for one
variable or another are available, including replacing $q$ in equation
\ref{eq:rkf-normerror} with its value subtracted from the system's
average velocity or center of mass for velocity or position
coordinates respectively, or adding a constant error softening value
to eliminate singularities in the error near zeros of the variable.
Following discussion in \citet{NumRec92}, one may also replace 
$|q|$ for some variables with its value added to its change at the 
last timestep:
\beq
q' = |q| + |\Delta t^n \dot q|.
\eeq
Alternate error metrics such as these have been implemented in VINE
with some success on several systems we have studied, however in
general, we expect that suitable error metrics will need to be worked
out on a case by case basis by the user. Some small comfort may be had
in the fact that, under most conditions, other error conditions are
more restrictive than the RKF error, making questions of the
suitability of the form of the RKF criterion moot.

\subsubsection{Differences in settings for time step criteria when
global or individual time steps are used, or for different problems}

For both of VINE's integrators, it is possible to check after each
time step whether the integration over that time step met or failed
the set of error criteria described above. If so, in principle one can
revert the time step and repeat it with a smaller step size. In
practice, reversion is only possible if the entire system is advanced
using a single, global time step for all particles, and is in fact
done in VINE when the global time step option is selected. When
individual time steps are used (discussed in section \ref{sec:ind_ts}
below), reverting the time step is usually not possible because it
requires keeping track of a large set of previous time steps for every
particle in the system, and is therefore usually prohibitive in terms
of memory as well as computational effort. Thus for an individual time
step scheme, the criteria for choosing the next time step and their
settings must be chosen more conservatively than with the global time
step option, in order to ensure in advance that the time step is small
enough to integrate that particle's properties correctly. For example,
\citet{bate1995} demonstrated that criteria similar to those in
equation \ref{eq:rkf-crit} give good results with the RKF integrator
when used alone with global timesteps, but that errors become
unacceptable when used alone with individual particle time steps.
Adding another criterion of the form of equation \ref{force_crit}
alleviated the problem. Similar situations may arise with the criteria
currently implemented in VINE when used on different problems. We have
therefore designed the error criteria code as a set of independent
routines for calculating specific error criteria, each with the same
interface and each called from a master routine, whose sole purpose is
to serve as a location at which criteria may be included or excluded.
The selection of which criteria to use and the addition of other
criteria can be done by the user with minimal difficulty, and the
timestep itself is computed from the minimum of all active criteria.

\subsection{The Individual Timestep Scheme}\label{sec:ind_ts}

In many astrophysical contexts, it is necessary to model the evolution
of regions with densities (or other quantities) that are orders of
magnitude larger than those of other regions in the same simulation,
or that change orders of magnitude more quickly. Such large variations
naturally introduce a wide range of physical timescales.  

Although it might be desirable in some cases to evolve all particles
with a single time step in order to maintain a highly stable
integration, the computational expense of doing so in all cases can be
prohibitive. Instead, it is possible to assign time steps for each
particle on an individual basis as required by a given particle and
thereby to evolve the particles independently. Assigning individual
time steps to the particles can speed up a simulation considerably,
since little processor time is wasted on evolving less dynamic regions
with the same small time steps as the most dynamic regions present in
the system.

Users of VINE can select whether to run simulations with either a
single global timestep or an individual time step for each particle.
Global time steps can be set either to the minimum absolute time step
(`global continuous' mode) or to the minimum binned time step ('global
binned' mode) corresponding to the time bins that would be used if the
individual time step option were active, as described below. The
latter option restricts the time steps to a discrete set of sizes and
therefore may enhance integrator stability, particularly for the
leapfrog integrator, in which fixed time steps are formally required
to retain the symplectic character of the integration. 

\begin{figure}[!t]
\epsscale{1.0} 
\plotone{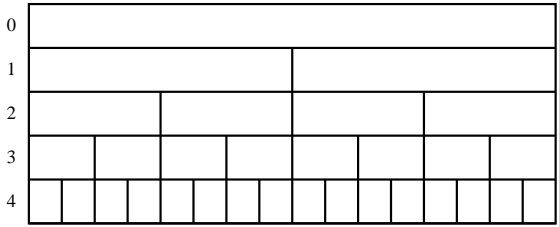}
\caption{\label{fig_tbins}
The individual time step scheme used in VINE. Only the five highest
levels of the time step hierarchy are shown here. Level 0 represents
the maximum time step, level 4 is a factor of $2^4$ smaller.}
\end{figure}

When the user selects individual time steps, VINE uses a variant of
the scheme proposed by \citet{hern_katz89} for their TreeSPH code,
based on previous work of \citet{porter85} and \citet{ewell88}. The
user first chooses a maximum time step allowed for the system, $\Delta
t_{max}$, which forms the top of the hierarchy of smaller time step
levels constructed by dividing the time step on the next higher level
by $2$ (see figure \ref{fig_tbins}). For time increments smaller than
the maximum, time is regarded as an integer quantity whose maximum
value is represented by a large power of $2$, in VINE set to $2^{28}$.
Each shorter time step can then be represented by a value $2^{28-n}$
where $n$ is the time step level counted from the top (see figure
\ref{fig_tbins}). Multiplying each integer by the smallest real valued
time step increment recovers the true, real valued time or time
increment, $\Delta t$, as needed. Time updates smaller than the
maximum are updated in integer increments, thereby avoiding errors
associated with time drift or finite precision truncation errors which
may occur when many small, real valued quantities are added together.
After the integration has proceeded through a full $\Delta t_{max}$
step, the real valued absolute time is incremented by $\Delta
t_{max}$.

VINE assigns each particle its own provisional time step as described
in \ref{sec_ts_crit}, then truncates it to the next smaller time step
level defined in the scheme. Time step assignment is done
transparently for both the leapfrog and RKF integrators using the same
binning scheme. Because the number of levels is finite, particles
evolve forward in groups corresponding to one or more levels of the
scheme that are currently active, rather than one by one on a
continuous spectrum of time steps. This is important for achieving
computational efficiency, both in serial and parallel operation,
because force computations for only one or a few particles at a time
are comparatively inefficient and difficult to balance among a large
array of processors running in parallel. Overheads associated with
repeatedly extrapolating all inactive particles are also minimized by
the grouping.

Three integer time variables are assigned to each particle, defining
the beginning of its current time step, $I_0$, its half time step,
$I_1$, and its time increment, $I_{dt}$. At the beginning of every
update, VINE performs a search through the list of all particles to
determine the closest future time for which any particle requires a
derivative calculation, defined by the condition that 
\beq
\label{eq:nexttime}
I_{\rm next} = \min_i(I_0 + I_{dt}, I_1)
\eeq
where the minimum is taken over all particles, $i$, and the resulting
value $I_{\rm next}$ defines the time at which the next derivative
calculation will occur. Using the value of $I_{\rm next}$, VINE sorts
particles onto three lists: particles for which $I_{\rm next}$ matches 
their end time step time, $I_0 + I_{dt}$, those for which $I_{\rm next}$
matches their half time step time $I_1$, and those for which neither
criterion applies. The three lists correspond to particles requiring
`full' updates to the end of their time step, those requiring `half'
updates at its midpoint, and those which do not require any update at
all. Simultaneously, VINE calculates a real valued time step
increment, $\Delta \tau$, for use during the integration for each
particle. For particles at either their half or full update step,
$\Delta \tau$ is identical to the particle's integration step, $\Delta
t$, otherwise, it defines the difference between the current time and
the time of its last update. The values of $\Delta \tau$ are then
given to the integrator, all particles are extrapolated to the current
time, a derivative calculation is performed and the system is
updated. Upon completion of the update, the values of $I_1$ for all
particles reaching their half step time are temporarily reset to
large values, in order to permit discovery of their end time step time
via equation \ref{eq:nexttime}.

The sequence of updates for different time step levels proceeds in
different order for the leapfrog and RKF integrators. For the RKF
integrator, particles on more than one time step level can be active
at the same time because a half update time at one level will
correspond exactly to a full update time at all lower levels. Time
therefore proceeds monotonically forward through the hierarchy. As
particles on finer levels complete their time steps, particles on
coarser levels join in and are integrated forward as well. In
contrast, because the leapfrog is an offset `DKD' integrator, particle
updates can occur for only a single time step level at a time because
half time steps on one level do not correspond to the times of half
steps on any other level. Also, when particles on coarser levels
become active, they advance forward by an increment larger than a full
time step on a fine level. This is important because it means that
inactive particles on coarse levels must be extrapolated both forward
and backwards in time at different points in the sequence as the
system is synchronized in preparation for the next derivative
calculation. No particle is ever extrapolated more than one half step
in either direction however, so the extrapolation remains within the
range spanned by the original update.


\section{Smoothed Particle Hydrodynamics}\label{sec:sph}

The Smoothed Particle Hydrodynamics (SPH) technique for simulating
hydrodynamic phenomena was first described by \citet{lucy77} and
\citet{ging_mon77}. Since then, much effort has gone into developing
the method \citep{ging_mon82,mon_ging83,monaghan85,monaghan88} and it
has become widely used in the study of many astrophysical problems
\citep[see][ for reviews]{benz88,benz90,monaghan92}.  

There are a variety of formulations of the SPH equations, different in
various details. In the following, we present the formulation
implemented in VINE, and will discuss differences from other
formulations and some of the corresponding advantages and
disadvantages separately in section \ref{sec:symmetrization}.

SPH solves the hydrodynamical equations in Lagrangian form and can be
regarded as an interpolation technique: the positions of the SPH
particles combined with an interpolation kernel define the fluid
quantities throughout the flow. By default, VINE implements the widely
used W4 smoothing kernel defined by \citet{mon_lat85} as:
\begin{equation}
\label{eq:kernel}
W(r_{ij},h_{ij}) = {\sigma \over h_{ij}^{\nu}} \left\{
\begin{array}{l@{\;\;}l}
1-{3 \over 2}v^2+{3 \over 4}v^3 & \mbox{if}\; 0 \leq v < 1,\\
{1 \over4}(2-v)^3 & \mbox{if}\; 1 \leq v < 2 ,\\
0 & \mbox{otherwise},\\
\end{array} \right. 
\end{equation}
where $\nu$ is the number of dimensions and $\sigma$ is the
normalization with values of $2/3$, $10/(7\pi)$ and $1/\pi$ in one,
two and three dimensions respectively. Fluid quantities at the
position of particle $i$ are then obtained as weighted sums over the
properties of all its neighboring particles. For the density, this
reads 
\begin{equation}
\label{rho}
\rho_{i} \equiv \rho(\m{r}_i) ~= \sum^{N}_{j=1} m_{j}~W(r_{ij}, h_{ij}) \,
\end{equation}
where $W$ is the kernel function defined by equation \ref{eq:kernel}
and $m_{j}$ are the masses of the $N$ neighboring particles. 

The dimensionless separation, $v=r_{ij}/h_{ij}$, between particles $i$
and $j$, requires the actual magnitude of the separation,
$r_{ij}=|\m{r}_i-\m{r}_j|$, and their characteristic `smoothing'
length scale, $h_{ij}$, defined as the mean of the smoothing lengths
of the two particles:
\beq
\label{eq:hij-defn}
h_{ij}=(h_{i}+h_{j})/2.
\eeq
Thus, with these definitions and the W4 kernel, particles whose
separations are $v<2$ contribute to the summations as `neighbors'.
Also, the influence of particles on each other is symmetric under
interchange of indices, an important characteristic required to ensure
conservation of momentum and other quantities. All other quantities
requiring symmetrization are defined similarly in VINE (see also
section \ref{sec:symmetrization}, below).

The choice of kernel used in VINE is made via an external module, so
users may substitute an alternate if desired with little or no change
to the rest of the code. The W4 kernel above is second order in the
interpolant and has the advantage of being defined on compact support:
particles more distant than $v=2$ do not contribute to the sum. As
written (see \vineII, section 3.5), VINE currently defines neighbors
as particles with $v<2$, consistent with the definition appropriate
for the W4 kernel. Other choices of kernel might include Gaussian
kernels used more commonly earlier in the history of SPH, which are
not defined on compact support at all, or other spline kernels,
perhaps compactly defined over a different range of separations, $v$.
For these kernels, a new neighbor definition must be made in the code,
to define an artificial cutoff or new separation range for the
neighbor search. This modification requires a change of only a single
line of code, and will therefore be trivial to implement. 


\subsection{The SPH formulation of the equations of hydrodynamics
implemented in VINE}\label{sec_sph_eq}

\subsubsection{Additions to the Momentum Equation}

When gas dynamics are included in a simulation, an additional term
must be included to equation \ref{eq:vel-deriv}, which models forces
due to pressure gradients and in its ideal form is:
\begin{equation}
\label{eq:mom}
\f{d\m{v}_{i}} {dt} = - {{\nabla p}\over{\rho}}
\end{equation}
where $p$ and $\rho$ are the pressure and mass density of the fluid,
respectively. This form defines the momentum equation appropriate for
`ideal' fluid dynamics. Explicit terms to model viscous processes
following the Navier-Stokes formalism are not included. The SPH
formulation used in VINE casts equation \ref{eq:mom} in the form:
\begin{equation}
\label{momSPH2}
\f{d\m{v}_{i}} {dt} = - \sum_{j} m_{j}
\left( \f{p_{i}}{\rho_{i}^2} + \f{p_{j}}{\rho_{j}^2} + \Pi_{ij}
\right) {\mbox{\boldmath$\nabla$}_{i} W(r_{ij}, h_{ij})} \,. 
\end{equation}
where $\mbox{\boldmath$\nabla$}_{i}$ means take the gradient with
respect to the coordinates of particle $i$ (see e.g.
\citealp{benz90}), $p$ is the gas pressure and $\rho$, the density, is
given by equation \ref{rho}. An additional term, $\Pi_{ij}$, appears
in equation \ref{momSPH2} but has no counterpart in equation
\ref{eq:mom}. Its purpose is described in section \ref{sec:avisc} and
takes the form of an artificial viscous pressure included to model
dissipative effects, without which there is no mechanism to convert
kinetic energy into heat due to non-reversible processes such as
shocks or viscosity. Alternatively (or in addition), one might include
a dissipative term modeled on the viscous terms found in the
Navier-Stokes equations directly, however, in its present form VINE
does not include such a term.

\subsubsection{The energy equation and equation of state}

The change in the thermodynamic state of the fluid requires an
evolution equation for a state variable corresponding to its internal
energy or entropy. \citet{springel2002} compare the benefits of
using one or the other formulation. VINE employs an equation for the
specific internal energy of the gas, defined for each particle. In the
simplest case of an ideal gas with no external heating or cooling
terms, only compressional heating and cooling are important and the
equation takes the form:
\beq
\label{eq:energy-eqn}
\f{du_{i}}{dt} = -{{p}\over{\rho}} \mbox{\boldmath$\nabla \cdot v$},
\eeq
where $\mbox{\boldmath$v$}$ is the local fluid velocity. Analogously
to the momentum equation above, the SPH formulation used in VINE casts
this equation with two terms as \citep{benz90}:
\addtolength{\arraycolsep}{-4pt}
\begin{eqnarray}
\label{idealenergy}
\f{du_{i}}{dt} &=& \f{p_{i}}{\rho_{i}^2} ~\sum^{N_{n}}_{j=1} m_{j} \,
\m{v}_{ij} \, \mbox{\boldmath$\cdot \, \nabla$}_{i} W(r_{ij}, h_{ij})
\nonumber \\
& &+ \f{1}{2} ~ \sum^{N_{n}}_{j=1} m_{j} \, \Pi_{ij} \, \m{v}_{ij} \,
\mbox{\boldmath$\cdot \, \nabla$}_{i} W(r_{ij}, h_{ij}) .
\end{eqnarray}
\addtolength{\arraycolsep}{4pt}
where the first term corresponds to reversible (`$PdV$') work. As
for the momentum equation, the second term including the viscous
pressure $\Pi_{ij}$ has no counterpart in equation \ref{eq:energy-eqn}.
It models the irreversible thermal energy generation produced by the
artificial viscosity as it attempts to model shocks or turbulent
energy dissipation. A master routine is available for users of VINE to
incorporate additional sources or sinks of internal energy via calls
to external routines, written to model a specific system. Switches to
activate calls to these routines are available to users in a text
input file, or can be trivially added to it, which is read in by VINE
at run time.

To close the set of equations, an equation of state must also be
defined to relate the internal energy, density and pressure to each
other. VINE includes options to call several simple equations of
state, such as isothermal, isentropic and ideal gases. While the
simple choices included will suffice for many problems, we recognize
that any set of equations of state will not be sufficient in general.
Therefore, as for energy sources and sinks, a master routine is
available for users to incorporate additional equations of state into
VINE with minimal difficulty through calls to external routines,
activated by switches set in a text input file.

\subsubsection{Artificial Viscosity}\label{sec:avisc}

VINE incorporates an artificial viscous pressure in its momentum and
energy equations to model irreversible thermodynamic dissipation from
shocks and viscosity. Following standard practice, the viscous
pressure includes both a bulk viscosity (the `$\alpha$' term) to
eliminate subsonic velocity oscillations and a von~Neumann-Richtmyer
viscosity (the `$\beta$' term) to convert kinetic energy to thermal
energy and prevent particle interpenetration in shocks
\citep{mon_ging83}. Formally, the expression for $\Pi_{ij}$ is 
\begin{equation}\label{eq:avform}
\Pi_{ij} = \cases{ 
(-\alpha_{ij} \, c_{ij}\,  \mu_{ij} + \beta_{ij} \,  \mu_{ij}^2)/\rho_{ij} &
$\m{v}_{ij} \mbox{\boldmath$\cdot$} \m{r}_{ij} \leq 0$, \cr 0
& $\m{v}_{ij}\mbox{\boldmath$\cdot$}\m{r}_{ij} > 0$, \cr } 
\end{equation}
where scalar quantities with both indices $i$ and $j$ are symmetrized
values using equation \ref{eq:sym-eqn} in section
\ref{sec:symmetrization} below. Vector quantities with two indices
represent differences, e.g. $\m{v}_{ij}=\m{v}_i-\m{v}_j$, and
$\mu_{ij}$ plays the role of the velocity divergence:
\begin{equation}
\label{bal_visc}
\mu_{ij} = \f{h_{ij} ~ \m{v}_{ij}\, \mbox{\boldmath$\cdot$}\,
\m{r}_{ij}}{\m{r}_{ij}^2 + \eta^2 h_{ij}^2} \; f_{ij} \,,
\end{equation}
with $\eta\approx 10^{-1} - 10^{-2}$ to prevent singularities. 

In its original form \citep{lmps86}, $f_{ij}$ is set to unity in equation
\ref{bal_visc}. \citet{balsara90,balsara95} found that this form gives
rise to large entropy generation in pure shear flows, which he
suppressed by defining an additional factor $f$ to reduce the
contribution selectively in such flow configurations. He defines $f_i$
as
\begin{equation}
\label{form_func}
f_i = \f{\left| \bigl< \mbox{\boldmath$\nabla \, \cdot \,$} \m{v}_i \bigr>
\right|}{\left| \bigl< \mbox{\boldmath$\nabla \, \cdot \,$} \m{v}_i \bigr>
\right| + \left| \bigl< \mbox{\boldmath$\nabla$} \, \m{\times} \, \m{v}_i
\bigr> \right| + \eta' }
\end{equation}
with $\eta'$ preventing divergence once more. VINE allows the user to
choose whether or not to allow the factor $f$ to vary or to set it
permanently to unity for all particles via a run time option.

The factors $\alpha$ and $\beta$ in equation \ref{eq:avform} are
parameters controlling the strength of the artificial viscosity. The
best choice for their values depend somewhat on the problem being
addressed in a particular simulation \citep[see e.g.][ for some common
choices]{stein_muell93,lsrs99}, but values near $\alpha=1$ and
$\beta=2$ are most commonly used in the literature. VINE allows users
to set both values in an input file or, at the user's option, to be
set dynamically as conditions warrant at different times or locations
in a simulation. The latter setting utilizes a variant of the
formulation designed by \citet{mor_mon97}, where each particle is
assigned time dependent viscous parameters, to minimize unwanted
viscous dissipation in quiescent flows while retaining good
reproduction of the flow properties in shocked regions where it is
required. In both standard test problems and in realistic cosmological
models of of galaxy clusters, \citet{dolag2005} found such a time
dependent AV formulation to be very useful, yielding much more
accurate results compared to the standard formulation with fixed
parameters. 

Following \citealt{mor_mon97}, each particle is assigned its own value
of the viscous coefficient, $\alpha_i$, which changes in time
according to a source and decay equation taking the form:
\beq
\label{eq:alpha}
{{d \alpha_i}\over{dt}} = -{{\alpha_i - \alpha_*}\over{\tau_i}} + S_i.
\eeq
The first term forces the value of $\alpha_i$ to decay asymptotically
to a value of $\alpha_*$ on a time scale $\tau_i$, given by
\beq
\label{eq:tau_av}
\tau_i = {{ \delta_\alpha h_i}\over{c_s\sqrt{(\gamma-1)/2\gamma}}},
\eeq
where $\delta_\alpha$ is a factor to relate the decay time scale to a
more convenient decay length scale describing the distance over
which the coefficient decays behind the shock. The form of the decay
time is derived by \citealt{mor_mon97} from the post shock Mach number
for strong shocks, combined with the sound speed and smoothing length. 

The second term in equation \ref{eq:alpha} is a source term of the
form discussed by \citet{rosswog00}:
\beq
\label{eq:s_term_av}
S_i = S_0 \max(-f_i(\alpha_{\rm max} - \alpha_i)
           \mbox{\boldmath$\nabla \, \cdot \,$} \m{v}_i , 0)
\eeq
where $\alpha_{\rm max}$ is a maximum value of the viscous coefficient
chosen to ensure that repeated strong compressions or shocks do not
increase the viscous parameter to values much larger than those known
to yield good results in test problems. The source function depends on
the velocity divergence, so that $\alpha_i$ grows in strong
compressions as required, and also includes the Balsara coefficient
from equation \ref{form_func} as suggested by \citealt{mor_mon97} in
order to suppress growth of the source function in shear flows. The
scale factor $S_0$ accounts for equations of state with $\gamma\not =
5/3$:
\beq
S_0 = [\ln({{5/3+1}\over{5/3-1}})]/[\ln({{\gamma+1}\over{\gamma-1}})],
\eeq
and is employed to ensure that the peak value of $\alpha$ remains
more nearly independent of the equation of state.

As implemented in VINE, each of the parameters $\alpha$, $\beta$,
$\alpha_{\rm max}$, $\alpha_*$ and $\delta_\alpha$ may be set at run
time by the user with appropriate choice made in a text input file.
Both the fixed and time dependent viscosity implementations retain the
von-Neumann Richtmyer term for each particle, with the ratio of
the two terms $\beta/\alpha$ also set by the user and fixed to the
same value for all particles. 

\subsection{Variable smoothing lengths}\label{sec:h_adapt}

With few exceptions, astrophysical systems exhibit moderate or large
density contrasts and ideally one would like to resolve these
contrasts as well as possible in simulations. The accuracy of the
interpolation of the SPH scheme depends on the number of neighboring
particles taken into account in sums like equation \ref{rho}, which
makes high numbers of neighbors (large $h$ values) desirable. On the
other hand, large neighbor counts increase the computational effort
and decrease the spatial resolution, since only scales larger than $h$
are resolved. These two competing effects lead to a requirement that
the number of neighbors should stay roughly constant at a level where
correct interpolation is assured without wasting computational
resources. Thus the particle smoothing lengths, $h$, need to vary in
space and time, i.e. each particle $i$ gets its own time-dependent
smoothing length, $h_i(t)$, which is then integrated according to:
\beq
\label{eq:hdot}
\f{dh_i}{dt} = - \f{1}{n_d} \f{h_i}{\rho_i} \f{d\rho_i}{dt} 
             = \f{1}{n_d} \, h_i \, \mbox{\boldmath$\nabla \, \cdot\,$}
\m{v}_{i} \;,
\eeq
where $n_d$ is the number of dimensions and the continuity equation
has been used to replace the time derivative of the density
\citep{benz90}. If the integration of equation \ref{eq:hdot} leads to
a neighbor count outside a given range, an exponential correction term
pushes $|\dot{h_i}|$ to greater values, such that more/less neighbors
are found on the next step if too few/many were found on the current
step. VINE attempts to keep the neighbor count within the range [30,
70] in 3D and [10, 30] in 2D. These ranges have been deliberately set
comparatively broadly because they depend somewhat on the total number
of particles \citep[see][ and references therein]{lsrs99} and on the
physics itself \citep{AGW07}. We recommend that in actual use, users
set somewhat more restrictive ranges, via parameters that are changed
when the code is compiled, and based on discussions in these
references.

The error introduced into the interpolation scheme by allowing for
varying smoothing lengths is second order in $h$
\citep{monaghan85,hern_katz89}. Since SPH with constant smoothing
lengths is also only accurate to second order in $h$ with the kernel
in equation \ref{eq:kernel} (see e.g. \citealp{benz90}), the addition
of variable smoothing lengths imposes no additional restriction.
However, variable smoothing lengths formally introduce an additional
term in the interpolated forms of the gradients of functions, as these
forms involve gradients of the kernel. These additional terms can be
neglected in many simulations (see \citealp{evrard88}), but doing so
may cause spurious entropy generation in some cases
\citealp{nel_pap94,springel2002,pm04}. In its present
form, VINE neglects the additional terms, but modifications to the
code to implement them via the `variational methods' derivations of
SPH discussed below are planned for the future.

\subsection{Symmetrization of quantities in SPH}
\label{sec:symmetrization}

The description of the SPH equations so far has only dealt with the
specific form implemented in VINE. Other formulations exist
\citep[e.g.][]{ging_mon82,evrard88,hern_katz89}, but a thorough review
of even the most common is beyond the scope of this paper. We suggest
\citet{thacker2000} and \citet{springel2002} for a more detailed
discussion of the alternatives, and limit ourselves here to an outline
of a few of the major points regarding the differences and why we use
the scheme implemented in VINE. 

An important difference between SPH implementations comes from the way
force symmetry between pairs of particles is handled. For momentum
conservation, symmetry between interchange of indices in the pairwise
force calculation between two particles is required. As noted above, 
VINE uses the arithmetic mean of quantities defined for each particle
to produce the symmetry:
\beq
\label{eq:sym-eqn}
q_{ij} ={{q_{i}+q_{j}}\over{2}},
\eeq
where $q$ is some quantity intrinsic to a particle such as smoothing
length or sound speed. Specifically for the calculation of the kernel
and its gradient, VINE uses a symmetrized smoothing length, $h_{ij}$,
so that the smoothing kernel defined for two particles is
\beq
W_{ij} = W(|\mathbf{r_i} - \mathbf{r_j}|, (h_i + h_j)/2).
\eeq
Alternatively, some implementations derive force symmetry from a
symmetrized kernel. Rather than defining $W(r_{ij},h_{ij})$, these
implementations define $\tilde{W}_{ij}$ as
\beq
\label{eq:kernsymm}
\tilde{W}_{ij}=\f{1}{2}W(|\mathbf{r_i} - \mathbf{r_j}|,h_i) +
               \f{1}{2}W(|\mathbf{r_i} - \mathbf{r_j}|,h_j),
\eeq
as originally suggested by \citet{hern_katz89} and used e.g. by
\citet{dave97,carraro98,lia2000,springel2001} and \citet{wadsley2004}.

The choice of kernel symmetrization has direct implications for the
neighbor search. To assure force symmetry, neighboring particles $i$
and $j$ must be determined to be mutual neighbors. Some codes use
range search techniques \citep[see e.g.][ note that only the serial
version of Gadget-1 uses this technique]{steinmetz96,springel2001}. In
this case, all particles $j$ with $|r_i-r_j| \le \eta \,h_i$ are used
as neighbors of $i$, with $\eta > 1$ and typically $\eta \approx 1.3$.
This method does not guarantee that both particles $i$ and $j$ will
find each other as neighbors. Instead, VINE uses a node opening
criterion based on the quantity $h_{ij}=(h_{i}+h_{j})/2$, which is
defined identically for all particle/particle pairs, particle/node
pairs and node/node pairs contained in the tree. It therefore assures
by construction that the mutual neighbor property is satisfied for all
particle pairs $i,j$. The higher computational cost (if any) required
to complete searches is more than offset by the fact that symmetrizing
the kernels through equation \ref{eq:kernsymm} leads to evaluating the
kernel function and its gradient twice, while VINE only needs to
evaluate them once.

A second important difference in symmetrization comes from differences
in the forms of the momentum and energy equations. Rather than
calculate a sum of two quantities, one from each particle in a pair, a
geometric mean might be used, or symmetry requirements relaxed
altogether and a non-symmetric form of the equations of motion
implemented. \citet{springel2002} compare these alternatives on
several test problems, along with two other formulations recasting the
energy equation instead as an entropy equation. They find that of the
three formulations using an energy equation, the best alternative is
an arithmetic mean. This result is, however, secondary in significance
to their finding that in cases where the system is under-resolved,
either spatially or temporally, {\it all} of the energy formulations
they studied may produce inaccurate results, as does one of their
entropy formulations. 

Instead, they recommend employing an entropy equation using a
symmetrization based on a novel variational methods derivation of the
equations of motion and of the entropy equation itself as formulated
in SPH. This form explicitly conserves entropy in adiabatic flows,
while avoiding the errors they demonstrated in the energy formulations
at low resolution. Roughly speaking, this formulation permits unequal
weights of the terms in the arithmetic means to be generated on a
pairwise basis depending on the details of the flow.
\citet{monaghan2002} also provides a variational methods based
derivation of the equations of motion, while retaining an energy
equation formulation, but does not specifically investigate its effect
on the outcome or accuracy of any test problems. 

When properly resolved, simulations performed with each of the five
formulations studied by \citet{springel2002} yield results consistent
with each other. Therefore, improvements derived from the variational
formalism will be of most value when resolution is limited, extending
the effective resolution of dynamical features in a simulation to much
lower particle counts. Users may properly be cautioned merely to
employ resolution in their simulations always sufficient to ensure
accurate results. Of course, such a requirement can rarely be met in
practice because of the highly problem dependent definition of
`sufficient resolution'. As \citet{nelson06} noted in discussions of
resolution requirements for disk simulations, formulations which
permit low resolution `failure' to be more graceful or more accurate
will be preferable to those whose failure modes produce larger errors
and inaccuracies. The fact that the variational formalism provides
such assurances is a point greatly in its favor and is a planned
addition to VINE.  At present however, a variational formulation of
SPH has not yet been incorporated.

\subsection{Additional time step criteria for SPH
simulations}\label{sec:tscrit-sph}

Two additional time step criteria are required in SPH simulations.
First the Courant-Friedrichs-Lewy condition must be satisfied:
\beq
\label{cfl}
\Delta t^{n+1}_{\mathrm{\mathrm{CFL}}} = \tau_{\mathrm{CFL}}\f{h_i}
              {c_i + 1.2\, (\alpha_i \, c_i + \beta_i \, h_i \max_j \mu_{ij})}
\eeq
where $\alpha_i$ and $\beta_i$ are the artificial viscosity parameters
and $c_i$ is the sound speed for particle $i$, and $\mu_{ij}$ is
defined by equation \ref{bal_visc} with the maximum taken over all
neighboring particles $j$ of particle $i$. The form of equation
\ref{cfl} is that suggested by \citet{monaghan89}, who recommend
values of $\tau_{\mathrm{CFL}} \approx 0.3$ for good results. 

Secondly, VINE requires that the smoothing length of each particle
must not change too much in one timestep:
\beq
\label{eq:hchange}
\Delta t^{n+1}_{h} = \tau_h \f{h_i}{\dot{h_i}} \;,
\eeq
where $\tau_h$ is a tuning parameter. We set $\tau_h \approx 0.1-0.15$
in order to ensure that particles encountering strong shocks require
several timesteps to pass entirely through the interface. For a gas
with a ratio of specific heats, $\gamma=5/3$ for example, the density
enhancement across a strong shock will be a factor of four,
corresponding to a smoothing length change of $4^{1/3}\approx 1.6$ in
3D. With a density jump of the severity and suddenness experienced in
a shock, the timestep restriction of equation \ref{eq:hchange} will
become important and the particle's timestep will decrease, allowing
better resolution of the physical conditions of the shock interface.
With $\tau_h=0.15$ for example, a particle will require at least
$\sim4-5$ timesteps to contract fully to the post-shock condition. 

The criterion in equation \ref{eq:hchange} is also beneficial in
situations where the spatial gradient of neighbors is large, as it
may be for particles on the surface of an object. Small changes in the
particle's smoothing length can then lead to very large changes in its
neighbor count and ultimately to large oscillations between far too
many and far too few neighbors, from one timestep to the next. A
timestep restriction dependent on the smoothing length variation
restricts such particles to correspondingly smaller timesteps, so that
such oscillations do not develop.

Preventing or damping such oscillations is important both because of
the comparatively high cost of computing the evolution of such
boundary particles, but also because they can lead to oscillations in
the physical quantities such as pressure forces, enhancing
oscillations further as particles experience large intermittent kicks.
Although certainly more fundamental in a physical sense, in a
numerical sense quantities such as density are actually derived from
numerically relevant smoothing length of a particle. Therefore, rather
than limiting the change in derived quantities such as density, VINE
limits the change to the particle's smoothing length, as a more direct
throttle on unphysical behavior. 


\section{Gravity}\label{sec:gravity}

The most costly calculation in nearly any particle simulation
including gravity is the computation of the gravitational forces of
the particles on each other. This is due to its long range nature,
which couples all particles in the system to each other. Modeling many
astrophysical systems requires forces accurate only to $\sim0.1-1$\%
and, if a system is collisionless, a force error of this magnitude
leads only to a marginal decrease of its collisional relaxation time,
therefore only minimally altering the overall evolution of the system
\citep[see e.g.][]{hhm93}. For such systems, an approximate solution
will be not only acceptable, but much to be desired if it can be
obtained much more quickly than an exact solution. For other models,
e.g. star clusters, highly accurate forces must be determined for
accurate evolution in spite of their cost; an approximate solution
will be useless. VINE implements both approximate and exact methods
for calculating mutual gravitational forces of particles on each
other. In this section, we provide an overview of our implementation
of these alternatives. We refer the reader to \vineII\  for a lower
level description of the detailed methods. Users can choose from among
the available options at run time an appropriate choice in a text
input file. 

\subsection{Exact gravitational forces obtained from direct
summation}\label{sec:direct-grav}

The most naive approach to calculating the magnitudes of interactions
of particles on each other is simply to calculate directly a sum of
the terms due to every particle on every other particle:
\begin{equation}\label{eq:gravity-n2}
\mbox{\boldmath$ a$}_{\rm i} = - \sum_j^{N_p} {{G m_j} 
           \over{|\mbox{\boldmath$ r$}_i - \mbox{\boldmath$ r$}_j}|^2}
            \m{\hat r}_{ij}
\end{equation}
where $\mbox{\boldmath$ r$}_i$ and $\mbox{\boldmath$ r$}_j$ are the
positions of particles $i$ and $j$, $\m{\hat r}_{ij}$ is the unit
vector connecting them and $m_j$ is the mass of the $j$'th particle.
This calculation is extremely expensive, requiring ${\mathcal
O}(N_p^2)$ operations for every time step. VINE includes two
alternatives for computing the sum in equation \ref{eq:gravity-n2} for
every particle. The sum can either be computed directly by the
processor, or it can be computed instead by special purpose hardware,
so called `GRAPE' coprocessors, described in section \ref{sec:grape}
below.

\subsection{Approximate gravitational forces obtained from tree based
sorting}\label{sec:tree-grav}

When physical models do not require or do not allow for exact
inter-particle gravitational force calculations, as will be the case
for so called `collisionless' systems, approximate forces may become a
desirable alternative if they speed up the required computations while
still retaining sufficient accuracy. For gravitational force
calculations, approximate forces can be obtained by aggregating the
contributions of distant particles into a single interaction, for
which all of their individual contributions together can be
approximated as being due to a sum of multipole moments, perhaps
truncated to some low order.

The difficulties in this approach lie in establishing how big each
aggregate can be before errors in the force become overwhelming, and
in sorting through all of the possible nodes, making certain that
every particle is included in the force calculation for every other
particle exactly once, either as an individual or as part of an
aggregate. The most common and most general purpose method for
obtaining such approximate forces is to organize the particles into a
tree data structure, and then use tree nodes as proxies for groups of
particles. By examining successive nodes in the tree, all of the
particles contained in that node can be either qualified as
interactors, or the node can be opened and its children examined for
acceptability instead. Fully traversing the tree produces a list of
nodes determined to be acceptable for interaction with a given
particle, and a list of atoms (single particles), for which an exact
computation is required. Using a tree to determine a list of neighbors
or acceptable nodes reduces the overall computational effort to
$\mathcal{O}(N_p \log N_p)$. 

The challenge in making the use of a tree efficient are first to
choose an efficient method to traverse the tree and, second, to choose
an efficient method to decide which tree nodes are acceptable as is,
and which must be separated into their constituent parts to be
examined in more detail. We will describe both the traversal strategy
and the node acceptability criteria used in VINE in detail in \vineII.
For purposes here, it will be sufficient to describe qualitatively the
criteria used to determine node acceptability.

In order to calculate an accurate gravitational force due to some node
which defines a particle distribution, the error it contributes to the
total force on a particle must be small. Mathematically, for a
multipole expansion to converge at all, this condition translates to
the physical statement that the particle on which the force acts must
be remote from the mass distribution defining the node \citep[see
e.g.][ ch. 4]{JacksonEM}. In addition, if the multipole expansion is
truncated rather than continuing to infinite order, as will be
desirable in an approximate calculation in order to save time, errors
corresponding to the neglected contributions from the higher order
terms must also be small.

VINE implements three runtime selectable options for determining the
acceptability of a node to be used in the gravitational force
calculation, each based on a different implementation of the
convergence radius of a multipole expansion and of limits on the
errors due to series truncation. Following common practice, we call
each of these options `Multipole Acceptance Criteria', or MACs. Each
MAC includes a user settable parameter `$\theta$' by which the
accuracy in different problems may be tuned to the most suitable value
for that problem, but the interpretation given to $\theta$ is specific
to each MAC. 

The first and simplest is the `geometric' MAC:
\begin{equation}\label{eq:geometric-mac}
r^2_{ij} > \left({{h_j}\over{\theta}} + h_i\right)^2 = R_{crit}^2,
\end{equation}
where $h_i$ and $h_j$ are the physical sizes of the particle and node,
respectively and $r^2_{ij}$ is the square of their separation. The
accuracy parameter $\theta$ takes a value between zero and one, and
parameterizes the minimum acceptable distance at which a node may be
used in the gravity calculation. Alternatively, by switching the
positions of $\theta$ and $R_{crit}$ and setting $h_i=0$ it may be
interpreted instead as the tangent of the angle subtended by the node
on the `sky' as seen by the particle. We incorporate the size of the
particle, $h_i$, into the MAC in order to ensure that the condition is
satisfied for all locations inside it, and to allow a generalization
of the MAC to be used for groups of particles taken together (see
\vineII\ for additional details). Errors due to truncation of the
multipole expansion are implicitly assumed to be small, because each
higher order term is diminished by an additional power of the
separation, $r$, appearing in the denominator of each multipole term.

Second, we implement the absolute error criterion of
\citet{salmon_warren94} (the 'SW' MAC), who derive explicit limits on
the errors due to series truncation. A node, $j$, is acceptable for use
in the gravity calculation if
\begin{equation}\label{eq:SW-mac}
r^2_{ij} > \left(h_i +
           {{h_j}\over{2}} +
        \sqrt{ {{h_j^2}\over{4}} +
              \sqrt{ {{3{\rm Tr}\, {\mbox{\boldmath$ Q_j$}}}\over{\theta}}}}
         \right)^2
\end{equation}
where $\theta$ is a value defining the maximum absolute error in the
acceleration that a single node may contribute to the sum and
${\mbox{\boldmath$ Q_j$}}$ is the quadrupole moment tensor for node
$j$. When the quadrupole moment is zero, this criterion reduces
exactly to the geometric MAC above with its $\theta$ defined to be
unity, defining a simple separation criterion. We note here also that
the original formulation of \citet{salmon_warren94} includes the
possibility of including the size of the particle, $h_i$, directly in
the term under the square root. Instead, we choose the form in equation
\ref{eq:SW-mac} in order to make possible a single calculation of the
MAC definition for each node prior to any tree traversals.

Finally, we implement the MAC discussed in \citet{springel2001}, which
we refer to as the `Gadget' MAC, and which uses an approximation for
the truncation error of the multipole expansion at hexadecapole order
to define an error criterion. A form using an octupole order error
formulation can be derived, but is computationally costly for modern
computers because it contains a square root, and is therefore not
used. For the relative error in the acceleration contribution of the
node, compared to the total acceleration at the last time step their
criterion is:
\begin{equation}\label{eq:gadget-mac}
r^6_{ij} > {{M_j h_j^4}\over{
                   \theta | \mbox{\boldmath$ a$}_{\rm old} |  }}
\end{equation}
where the gravitational constant implicitly present in the formula is
set to $G=1$, $\mbox{\boldmath$ a$}_{\rm old}$ is the value of the
acceleration for a particle at the last time it was calculated and
$\theta$ is a dimensionless maximum relative error in the acceleration
to be allowed to any acceptable node. Because it requires a previous
value of the acceleration, equation \ref{eq:gadget-mac} cannot be used
for the first calculation. Instead, we use the geometric MAC of
equation \ref{eq:geometric-mac} with its accuracy parameter set to
$\theta=0.5$. Also, equation \ref{eq:gadget-mac} may not ensure that
particle and node do not overlap in space, violating the separation
condition required for convergence of the multipole summation.
Therefore, when the Gadget MAC is selected, we also require that
equation \ref{eq:geometric-mac}, with its parameter set to $\theta=1$,
is satisfied.

\subsubsection{The acceleration calculation}\label{sec:treegrav-acc}

To compute the gravitational acceleration due to each entry on the
list of acceptable nodes on some particle, $i$, users of VINE may
choose one of two options. First, they may choose to sum the
contributions from acceptable nodes and atoms using a multipole
summation truncated at quadrupole order and computed on their general
purpose `host' processor or, alternatively, they may instead choose to
compute the summations to monopole order using a GRAPE coprocessor, if
one is available.

For the former, VINE sums the multipole series for each acceptable
node using the unreduced quadrupole moment formulation described in
\citet{bbcp90}. Mathematically, the acceleration on particle $i$ due
to node $j$ is: 
\begin{eqnarray*}
\label{eq:pole-accel}
\mbox{\boldmath$ a$}_{\rm i} & = & M_j f(r)\mbox{\boldmath$r$}
     + {{f'(r)}\over{r}} \mbox{\boldmath$Q$}_j\cdot \mbox{\boldmath$r$} \\
      && + {{1}\over{2}}\left[
          {{f''(r)}\over{r^2}}
             \mbox{\boldmath$r$} \cdot \mbox{\boldmath$Q$}_j \cdot
                                           \mbox{\boldmath$r$} \right.   \\
      && + \left. {{f'(r)}\over{r}} \left({\rm Tr\,}\mbox{\boldmath$Q$}_j
           - {{\mbox{\boldmath$ r$} \cdot \mbox{\boldmath$ Q$}_j \cdot
                      \mbox{\boldmath$ r$}}\over{r^2}}\right)
       \right]\mbox{\boldmath$ r$}  \\
\end{eqnarray*}
where the function $f(r)=Gr^{-3}$, \mbox{\boldmath$ r$}=
$\mbox{\boldmath$ r$}_j - \mbox{\boldmath$ r$}_i$ and
$\mbox{\boldmath$ Q$}_j$ is the quadrupole moment tensor for node $j$.
No specific gravitational softening (see section \ref{sec:softening})
is required in equation \ref{eq:pole-accel} because the criteria used
to determine the list of atoms ensures that no nodes will require
softening. Atoms are handled independently and contribute only a
monopole term, softened according to the options discussed below. The
acceleration due to all atoms and nodes together are summed to obtain
the gravitational acceleration acting on a particle, i.e. the right
hand side of equation \ref{eq:vel-deriv}. 

\subsection{GRAPE hardware}\label{sec:grape}

As noted above, VINE includes options to calculate gravitational
forces using GRAPE coprocessors if they are available, either in
approximation or exactly. `GRAPE' hardware (for
\textbf{G}r\textbf{A}vity \textbf{P}ipelin\textbf{E}) to accelerate
$N$-body calculations has been developed by several collaborations in
Japan \citep[][ see also \citealp{makino_taiji98} for a
review]{sugimoto90, fimesu91, ito91, okumura93, mtes97, kawai2000,
makino2003, fukushige2005}, and has been used both by them and by many
others throughout the astronomical community \citep[see, e.g.,
][]{naab2003, athan2003, naab2006, merritt2006, portegies2006,
berczik2006}. This hardware is attractive to users because it is a
system of specialized computer chips designed to perform a summation
of $1/r^2$ force calculations very quickly using a highly parallelized
pipeline architecture. These chips are combined onto a processor board
which also hosts all other other necessary functional units, including
memory, I/O controller, etc, and which is then attached to a host
computer, usually a desktop workstation. which performs the rest of
the calculations required in a simulation.

Two types of GRAPE boards exist, differentiated by even or odd version
numbers. GRAPE boards with odd numbers have a less accurate numeric
format, so that the relative error of the total force on a particle is
of the order of $2\%$ for GRAPE-3 \citep{okumura93} and $0.3\%$ for
GRAPE-5 \citep{kawai2000}. For collisionless systems, this imposes no
problem for the time evolution, as the errors are uncorrelated
\citep{mie90}. GRAPE systems with even version numbers, such as
GRAPE-2 \citep{ito91}, GRAPE-4 \citep{mtes97} and the most recent,
GRAPE-6 \citep{makino2003} have higher accuracy in their internal
numeric formats and were designed for simulating of collisional
systems like globular clusters. 

VINE supports GRAPE-3, GRAPE-5 and GRAPE-6 boards, the latter in both
a full and reduced size `MicroGRAPE' form, also known as 'GRAPE-6A'
\citep{fukushige2005}. The code can use GRAPE boards for direct
summation of forces as well as a component in an approximate tree
based gravity calculation. We describe the details of this approach in
\vineII\  after the tree itself has been described in detail.

\subsection{Softening the forces and potential}\label{sec:softening}

Simulations of physical systems depend on a choice to assume that the
system is either `collisional' or `collisionless', corresponding to
the statement that the evolution of the system as a whole is or is not
strongly affected by the outcomes of interactions between individual
particles (`collisions'). In hydrodynamic systems for example,
collisions between individual atoms or molecules matter only in the
aggregate, as a pressure. In $N$-body systems of galaxies for another
example, where individual particles may in fact be stars rather than
atoms, a similar statement can be made if the two body relaxation time
is long compared to the expected simulation time or the lifetime of
the system.

Even though the underlying physical systems may be collisionless in
the sense that no particular interaction affects the result,
simulations of them may not be because the number of particles used in
the simulation is typically many orders of magnitude smaller than the
actual number of bodies in the real system. Particles in these
simulations are actually meant to represent an aggregate of some very
large number of physical particles. Actual collisions between them, as
individuals, are therefore unphysical. In order to recover the
collisionlessness of the physical system, forces between particles
must be `softened' in some manner or, in other words, the particles
must be provided with an internal density distribution consistent with
their interpretations as aggregates of many particles rather than as a
distinct entity.

In practice, softening is achieved by modifying the gravitational
potential on small scales to avoid the pure $1/r$ form and the
associated numerically infinite forces at small separations.  The
softened form actually defines a mass density distribution for each
particle, so that they include an assumption of some spatial extent,
rather than that they are point like objects. Using softening, the
artificial close encounters between $N$-body particles are suppressed
and the particles evolve under the action of the global gravitational
potential created by all particles. 

\subsubsection{Forms of gravitational softening in
VINE}\label{sec:soft-form}

A long standing debate pervades the astrophysical literature
concerning how gravitational softening should be implemented in
particle simulations \citep[see e.g.][and references therein for
detailed
discussions]{merritt96,bate_burkert97,romeo97,afl2000,dehnen2001,nelson06}.
No form of softening so far proposed exists which is free of defects
and some care must be taken to ensure simulation results do not depend
on the implementation of softening. In practice, two alternatives are
common; first to use `Plummer softening' and second to use a spline
based kernel, perhaps also with a space and time dependent length
scale. VINE implements both options, via a user selectable switch. 

Plummer softening defines the density function of each particle
to be a Plummer sphere, so that the force on particle $i$ due to
particle $j$, at a distance $r_{ij} = |\m{r}_{i} - \m{r}_{j}|$, 
becomes 
\begin{equation}\label{eq:Plummer-force}
\m{F}_i = -{{G m_i m_j}\over{r^2 + \epsilon^2}} \m{\hat r}_{ij},
\end{equation}
where $\m{\hat r}_{ij}$ is the unit vector connecting particles $i$
and $j$. The potential is defined similarly as 
\begin{equation}\label{eq:Plummer-pot}
\Phi = -{{G m_j}\over{(r^2_{ij} + \epsilon^2)^{1/2}}},
\end{equation}
where $\epsilon$ is a softening length scale. Plummer softening was
originally motivated by \citet{aarseth1963}, who used it in
simulations of clusters of galaxies. Its most compelling advantage is
that it is simple to implement and computationally inexpensive. Its
major drawback is that it never converges to the exact Newtonian
potential at any separation.

Spline based softening defines the density function of a single
particle using a predefined kernel that extends over some finite
region in a manner not unlike that used to derive hydrodynamic
quantities in SPH. Although in principle any kernel could be used,
VINE implements the kernel defined in equation \ref{eq:kernel}, used
for the SPH interpolations. This kernel has the advantages that it has
compact support, i.e. for $r>2h$ it takes the exact Newtonian form,
that for hydrodynamic simulations with smoothing and softening set
equal, pressure and gravitational forces are nearly equal and opposite
on small scales \citep{bate_burkert97}, and that for a given number of
particles, errors in the force calculation are smaller than those of
many other possible kernels \citep{dehnen2001}. 

To specify the force on particle $i$ due to particle $j$, we apply
Gauss's law and integrate over the kernel density distribution to
obtain the fraction of the source particle's ($j$) mass enclosed by a
sphere with radius equal to the separation between them:
\begin{eqnarray}\label{eq:massgrav}
\hat m(r_{ij}) & = & 4 \pi \int_0^{r_{ij}} \rho(v) v^2 dv \nonumber \\
               & = & 4 \pi \int_0^{r_{ij}} W(v,h_{ij}) v^2 dv \, ,
\end{eqnarray}
where the density, $\rho$, is replaced by the softening kernel $W$ in
the second equation, and $v$ is as defined in section \ref{sec:sph}.
In 2D, a similar form for $\hat m_j$ can be defined but leads to a
finite, {\it non}-zero force at zero separation because the conceptual
framework that underlies the definition does not carry over unaltered
from three dimensions to two \citep{nelson06}. VINE instead modifies
the form of the kernel in 2D to that proposed by \citealt{nelson06},
to avoid the inconsistency. Given the modified definition of the
source particle's mass, the force and potential are defined by the
equations for Newtonian gravity: 
\begin{equation}\label{eq:spline-force}
\m{F}_i = -{{G m_i \hat m_j}\over{r^2_{ij}}}\m{\hat r}_{ij}
\end{equation}
and
\begin{equation}\label{eq:spline-pot}
\Phi = -{{G \hat m_j}\over{r_{ij}}}.
\end{equation}
Note that as two particles come closer (without decreasing their
softening lengths), the force decreases to zero because the mass
enclosed decreases to zero, and that the force is anti-symmetric with
respect to interchange of $i$ and $j$. At great distances, $\hat m_j$
becomes $m_j$ and the exact Newtonian form is recovered. 

Three softening variants are available in VINE, via options specified
at runtime. First, Plummer softening may be selected with a fixed
softening length $\epsilon$ for all particles. When the GRAPE option
(see section \ref{sec:direct-grav}) is selected for gravitational
force calculation, this is the only option available due to hardware
constraints. Second, either `fixed' or `variable' kernel softening may
be selected, with the latter option affecting only the treatment of
SPH particles. If the fixed option is selected, each SPH particle $i$
is softened using a single softening length $h_i=\epsilon$, specified
at run time for all particles. After the gravity calculations are
completed, smoothing lengths are reset to their locally defined,
spatially and temporally variable values so that later hydrodynamic
calculations may proceed. If the variable option is selected, the
individual (and time varying) smoothing lengths of each particle are
used as individual softening lengths. In both the fixed and variable
kernel softening options, all $N$-body particles use the kernel and
their predefined (fixed) softening lengths $h_i$. This alternative
allows several species of $N$-body particles to be included, each with
their own (possibly different) softening length. 

Depending on the softening selection, either terms from equation
\ref{eq:Plummer-force} or equation \ref{eq:spline-force} (in the form
of accelerations) are summed over all particles derived from the tree
traversal to obtain the gravitational acceleration acting on a
particle, to be in the right hand side of equation \ref{eq:vel-deriv}.

\subsubsection{Spline softening and the connection with
SPH}\label{sec_gravneigh}

Calculation of gravitational forces between neighboring SPH particles
requires the identification of equation \ref{eq:kernel} as a density
distribution, a different and stronger assumption that its use as an
interpolation kernel in non-self gravitating SPH. Two important
consequences arise from the identification. First, although there is
no requirement that the hydrodynamical smoothing length and the
gravitational softening lengths are equal, large force imbalances may
develop if they are not, due to the different assumptions about the
mass distribution within a single particle. The imbalances require
careful consideration because they are consequences of the numerical
assumptions, not of the physical systems, but can catastrophically
change the outcome of a simulation. We refer the reader to
\citet{bate_burkert97} and \citet{nelson06} for more detailed
discussions, but note the main conclusion of both is that more
favorable outcomes are obtained when gravitational softening and
hydrodynamic smoothing lengths are set to be equal. 

Secondly, equation \ref{eq:massgrav} will introduce fluctuations into
the potential energy of the system (and of individual particles) when
it is used with variable smoothing lengths because the change in mass
distribution within a particle is not accounted for in the potential
when its smoothing length changes. In most cases, the variation will
be small because it comes only from SPH neighbor particles, rather
than from the entire mass distribution. However, in some cases it may
be an important local or global effect, such as in uniformly
collapsing or expanding media, or in cases when a small subregion
collapses by many orders of magnitude. This violation of energy
conservation due to smoothing length variations can be estimated as
\citep{benz90}: 
\begin{equation}\label{eq:h_adapt_pot_change}
dH = -G \sum_{i<j} m_{i} m_{j}
          \f{\partial \Phi}{\partial h_{ij}}    dh_{ij}.
\end{equation}
In principle this formula could be used to correct the potential
energy of SPH particles (but not their accelerations), but we have not
included it in VINE because it requires an additional search for
neighbor particles, with its associated computational cost.  We also
note that a conservative form of softening has recently been developed
by \citet{pm07}, but this form has yet to be implemented in VINE.


\section{A third particle species: `star' particles}\label{sec:PM}

Astrophysical systems frequently include multiple physical components
with substantially different characteristics, embedded within and
interacting with each other. Some examples of such components can be
satisfactory modeled with the two particle types (SPH and $N$-body
particles) already discussed. For example, the example simulation in
section \ref{sec:set-params} includes an interstellar gas modeled with
SPH particles as well as three additional components (stellar disk,
stellar bulge, and dark matter halo) each modeled with $N$-body
particles. In addition to the possibility that these sorts of fluid
and non-fluid components could interact, another common characteristic
in astrophysical systems is the interaction between a comparatively
low density gaseous medium and one or more massive objects, such as
stars. Such systems include, for example, a single star surrounded by
a gaseous circumstellar disk \citep{NBAA,NBR}, or cloud of gas in
which one or many stars form during the evolution \citep{BCB06}. For
modeling simplicity, each component may require separate treatment in
order to capture essential physical phenomena peculiar to only one
without distorting the evolution of another, and include explicit
coupling terms to model their interactions.

In addition to the capability to evolve systems simulated in the
purely $N$-body and SPH particle frameworks already described, VINE
also includes the capability to model a third species of particles,
similar to $N$-body particles, but with additional characteristics not
present in the other types. This species is designed to be used in
models requiring a small number of stars or other similar point masses
in astrophysical systems, and interact with the rest of the system as
gravitational `sinks'. VINE evolves star particles using the same
integration method used for all other particles (section
\ref{sec:integ}), and with the same time step criteria, though
different coefficients may be assigned. They interact gravitationally
with all other particles and with each other according to a softened
Newtonian force law, as described in section \ref{sec:softening}.
Because they may be many orders of magnitude more massive than the
other types of particles, they are not included in the approximate,
tree based force calculation, but instead take advantage of the fact
that there will typically be a very small number of such particles to
obtain forces via direct summation.

The principle difference between VINE's implementation of star
particles and the simple $N$-body particles already discussed is their
ability to `accrete' particles of the other two types, should their
trajectories bring them into close proximity. If activated by a switch
set by the user at run time, VINE makes a check to determine whether
any particles have moved to within one smoothing length, $h_*$, of
each star after each timestep. When individual time steps are used,
both the accretor and accreted particles must be at the end of their
time step, in order to maintain simplicity in the code and the
integration scheme. When a particle is found, it is removed from the
simulation and its mass and momentum are added to the star. In order
to conserve the center of mass of the system, the star particle is
artificially moved to the common center of mass of the pair.
Similarly, VINE calculates the angular momentum of the star and fluid
particle around their common center of mass, then adds it to the star
as an internal `spin', in order for any later accounting of the
system's total angular momentum to be conserved.


\section{Boundary Conditions}\label{sec:bounds}

In many contexts, simulations of entire physical systems may be
possible for many astrophysical systems of interest. Other systems may
be studied only as some small subset of a larger whole, e.g., a
star formation simulation which models only a part of the parent
molecular cloud. In either case, a complete model requires that
conditions at the boundaries of the computational domain be specified
in order to model the influence of matter from outside it. 

For particle based simulations of entire systems, `free' boundary
conditions require no special treatments because the fluxes of
material, momentum and energy are carried by the particles themselves.
Particles feel perturbations only from other particles already present
in the simulation, not from any specific boundary, and move into
adjacent to empty regions as conditions warrant. Otherwise, such
regions require no computational effort. More complicated systems
require active management of particles near boundaries, with specific
treatments tailored to specific problems.

For the example star formation simulations above, the relevant treatment
will be of a volume of space surrounded by an infinite series of
other, identical volumes, replicated in succession at greater and
greater distances in each direction from the original. VINE includes a
module to implement periodic boundary conditions in this context and
we describe it below. Being much rarer in astrophysical contexts, we
have chosen not to implement reflecting boundaries, for which material
approaching some surface is repelled with identical velocity but
opposite sign, although a template routine has been included for this
purpose. 

\subsection{Periodic Boundaries}\label{sec:periodic-bc}

When periodic boundary conditions are employed, all matter resides
inside a predefined simulation box. The boundary conditions then imply
that calculations of physical quantities, such as gravitational
forces, gas densities, etc. are carried out as if the box were
surrounded in every direction by a series of identical replicas,
extending to infinite distance in all directions. Particles whose
trajectories pass through one of the box's boundaries, entering one of
the replications, are artificially restored to the original box on the
opposite boundary. Therefore, after every position update, VINE
performs a check for particles which have moved outside of the box. If
any such particles are detected, VINE adds or subtracts the length of
the box to the appropriate position component of those particles,
effectively reinserting them on the opposite side of the volume from
the one from which they exited. Velocities are unchanged. 

\subsubsection{Gravitational Periodic Boundaries}

Various treatments of gravitational periodic boundary conditions in
particle evolution codes have been discussed in the literature. Most
are based on the Ewald method \citep{ewald1921} and we refer readers
to \citet{hbs91} for the application to astrophysical particle
simulations\footnote{note however that their eq. 2.14b is missing a
factor $|\m{r} - \m{n}L|$, see \citealp{klessen97}.}. Our
implementation closely follows \citet{klessen97}, so here we describe
only the overall approach and basic equations and refer the reader to
his work for additional details regarding implementation, accuracy and
performance.

When a simulation uses periodic boundaries, the Newtonian potential of
the particles must be replaced by the potential of an infinite,
periodic replication of those particles. Thus the Poisson equation
becomes
\beqa
\mbox{\boldmath$\nabla$}^2 \phi(\m{r}) & = & 4 \pi G \rho(\m{r})\nonumber\\
& = & 4 \pi G \sum_\m{n}\sum_{j=1}^N m_j \delta(\m{r}-\m{r}_j-\m{n}L)
\eeqa
where $L$ is the period and $\m{n}$ an integer vector. The general
solution can be written as 
\beqa
\label{poissoneq}
\phi(\m{r})&=&-G\sum_\m{n}\sum_{j=1}^Nm_j\mathcal{G}(\m{r}-\m{r}_j-\m{n}L)\\
&=&\f{G}{L^3}\sum_{j=1}^N\sum_\m{k}\hat{\mathcal{G}}(\m{k})m_jj
  \exp\left[i\m{k}(\m{r}-\m{r}_j \right]
\eeqa
with the Green's function for the gravitational potential given by
$\mathcal{G}(\m{r})=1/r$ or, in Fourier space, $\hat{\mathcal{G}}=4
\pi/k^2$. The sum in equation \ref{poissoneq} converges only slowly.
Ewald's method alleviates this problem by splitting the summation into
a short range part, which converges quickly in real space and a long
range part, which converges quickly in Fourier space. So the Green's
function becomes $\mathcal{G}=\mathcal{G}_S+\mathcal{G}_L$ with
\beqa
\mathcal{G}_S&=&\f{1}{r} \erfc(\alpha r); \ \ \ \ 
         \hat{\mathcal{G}}_S(\m{k})=\f{4 \pi}{k^2}\left[ 1-\exp\left(-\f{k^2}{4 \alpha^2} \right)\right]\\
\mathcal{G}_L&=&\f{1}{r} \erf(\alpha r); \ \ \ \ \ 
          \hat{\mathcal{G}}_L(\m{k})=\f{4
  \pi}{k^2} \exp\left(-\f{k^2}{4 \alpha^2} \right)
\eeqa
where $\alpha$ is a scaling factor with units of an inverse length,
$\erf(x)$ and $\erfc(x)$ are the error function and its complement,
respectively. To balance good accuracy with low computational cost,
the parameter values $\alpha=2L$, $|\m{r}-\m{n}L|<3.6L$ and
$\m{k}=2\pi\m{h}/L$ with $\m{h}$ being an integer vector with
$|\m{h}|<10$ are a reasonable choice \citep[see e.g.][]{hbs91}.
The resulting equation for the potential is
\beqa
\phi(\m{r})=&-&G\sum_{j=1}^Nm_j \left[\sum_\m{n}\f{\erfc(\alpha
    |\m{r}-\m{r}_j -\m{n}L|)}{|\m{r}-\m{r}_j -\m{n}|}\right.\nonumber\\
 &-&\left.\f{1}{L^3}\sum_k\f{4\pi}{k^2} \exp\left(-\f{k^2}{4\alpha^2}\right)
     \cos \left[\m{k}(\m{r}-\m{r}_j\right]\right]
\eeqa
and the force on a particle $i$ is
\beq
\m{F}(\m{r}_i)=-m_i\mbox{\boldmath$\nabla$}\phi(\m{r}_i)=Gm_i\sum_{j\ne
i}\m{f}(\m{r}_i-\m{r}_j)
\eeq
with the function $\m{f}$ defined as 
\beqa
\label{pbforceeq}
\m{f}(\m{r})=&-&\sum_\m{n}\f{\m{r}-\m{n}L}{|\m{r}-\m{n}L|^3}\left[\erfc(
    \alpha|\m{r}-\m{n}L|) \right.\nonumber\\
&+&\f{2\alpha}{\sqrt{\pi}}\left. |\m{r}-\m{n}L|\exp(-\alpha^2|\m{r}-\m{n}L|^2)
    \right]\nonumber\\
&-& \f{1}{L^3}\sum_\m{k}\f{4\pi\m{k}}{k^2} \exp
  \left(-\f{k^2}{4\alpha^2} \right) \sin(\m{kr}) 
\eeqa

Most practical implementations use a modified force law for computing
the forces from any node or particle on the interaction list obtained
from traversing a tree \citep[see e.g.][]{dave97}. When GRAPE hardware
is used however, no modified force law can be used, since only an
unmodified Plummer force is programmed into the hardware itself. In
order to preserve the possibility of the use of GRAPE hardware in VINE
in this case as well, we have therefore chosen an alternate
implementation of Ewald's method \citep{klessen97}. Instead of
modifying the force law for all interactions inside the simulation
box, the forces from matter inside the simulation box are computed
as if no boundary was present, then all particles get a correction
force to account for the contributions from neighboring boxes. As in
equation \ref{pbforceeq} above, all pairwise interactions between
particle pairs are included and we only want to compute the correction
due to all $L$-periodic replications outside of the box, we simply
need to subtract the interaction of the pair inside the box:
\beq
\m{f}_\mathrm{cor}(\m{r})=\m{f}(\m{r})+\f{\m{r}}{|\m{r}|^3}
\eeq
Similarly the pure correction for the potential is
$\phi_\mathrm{cor}(\m{r})=\phi(\m{r})+1/|\m{r}|$. Note that
$\m{f}_\mathrm{cor}$ and $\phi_\mathrm{cor}$ actually represent the
corrective Green's functions for force and potential, respectively. 

To determine the correction, VINE maps the particles onto a grid using
the CIC (Cloud-In-Cell) scheme \citep[see e.g.][]{hockney81}, creating
a density distribution on the grid, which is then Fourier transformed.
The result is convolved with the Fourier transform of the Green's
function, $\hat{\m{f}}_\mathrm{cor}$, describing the
correction and then transformed back into real space. Finally, the
forces and potential are mapped from the grid back onto the particles,
giving every particle the desired correction of the force and
potential at the particle's location. Tables with the Fourier
transformed correction terms are pre-computed, so that during the
simulation only the transform of the density field needs to be
calculated, then it is convolved with the corrections and transformed
back. For a detailed description of the method, we refer the reader to
\citet{klessen97}. 

The computational cost of the calculation of the correction terms is
governed first by the grid resolution and second by the speed of the
FFT algorithm used. \citet{klessen97} shows that sufficiently accurate
forces can be obtained if the linear density of grid cells is at least
twice that of the particles. For example, if e.g. $64^3$ particles are
used, the grid should have a resolution of at least $>128^3$. 
 
Finally, if the periodic boundary conditions are used in connection
with the individual particle timestep scheme (see section
\ref{sec:ind_ts}), the periodic correction part of the force does not
have to be calculated on every particle timestep, which would
otherwise be a significant overhead. Instead, a CFL-like condition can
be applied to calculate new corrections whenever a particle could have
traveled a given fraction of a grid cell since the last time the
periodic corrections have been calculated.

Most publicly available and vendor supplied numerical libraries
include some variant of fast, parallel FFT algorithms. Rather than
writing multiple variants of our Ewald code for each of these
libraries, we link to the FFTW library \citep{fftw05}, commonly
available on most computing platforms, in order to retain both maximum
portability for VINE and to realize high performance on all platforms.

\subsubsection{SPH Periodic Boundaries}

Calculations of hydrodynamic quantities are always local in the sense
that only neighboring particles contribute to any given particle's
hydrodynamic quantities. No complex algorithms, such as Ewald's method
for gravity, are required to determine forces due to particles in
neighboring replications of the simulation box. The only change that
must still be made is to modify the definition of the distance between
pairs of particles, so that particles located near a boundary see
neighboring particles from both sides of that boundary.

The required modification utilizes the fact that no particle can be
more distant from another than one half of the box size in any
direction. If it were, its ghost particle in a neighboring box would
instead be chosen as neighbor because its separation in that direction
was smaller. We can therefore define the separation in each direction
as
\begin{equation}\label{eq:periodic-sep}
\delta x_{ij} = x_i  - x_j - L\ {\rm nint}\left( {{x_i - x_j}\over{L}}\right)
\end{equation}
where $L$ is the box length and `{\tt nint}' stands for the `nearest
integer' function and takes the value of zero or $\pm1$, depending on
the value of its argument. With this definition, and accounting for
the periodicity of the box, the separation of a pair of particles can
be calculated simply as the difference between their natural
coordinates, plus an extra term which reduces to zero when the natural
separation is small. If the magnitude of their natural separation is
larger than $L/2$, the box length will be added or subtracted as
appropriate. Combining the separations in each of the coordinate
directions yields a distance. The computational cost of the extra
operations are quite small, and no additional code infrastructure is
required to handle the existence and storage of actual ghost
particles. No additional infrastructure is required for neighbor
searches either, since equation \ref{eq:periodic-sep} is trivially
generalizable to calculations involving tree nodes as well. After
neighbor identification and distance calculations are complete,
computations of all hydrodynamic quantities, whether involving the
distance metric directly or indirectly, through the SPH kernel (e.g.
densities or velocity and pressure gradients), proceed as in the case
of free boundaries.

The definition of separation in equation \ref{eq:periodic-sep} has one
minor side effect, that any particle so large that it is a neighbor of
both some other particle contained in the simulation volume {\it and}
one or more of that particle's duplicates in neighboring volumes will
find only one instance of that particle. Its duplicates will not be
found. We consider this situation to be extremely unlikely for any
simulation containing more than a few tens of particles, and so
neglect the possibility entirely.

\section{Test Simulations}\label{sec_tests}

In this section we present various tests of the code, demonstrating
the capabilities of VINE for SPH simulations as well as for 
$N$-body simulations. 

\subsection{Adiabatic collapse of a cold gas
sphere}\label{sec:sphere-test}

Since \citet{evrard88} the adiabatic collapse of a cold, initially
isothermal gas sphere under its own gravity has been a widely used
test case for SPH codes, see e.g. \citet{hern_katz89,stein_muell93,
hult_kaell97,carraro98,springel2001,thacker2000}. Adopting a similar
setup as these authors, we simulate the collapse of a spherically
symmetric gas cloud with density profile
\beq
\rho(r)=\f{M}{2 \pi R^2} \f{1}{r}
\eeq 
where $M$ is the total mass of the cloud and $R$ is its maximum
radius. For simplicity we choose a unit system with $G=M=R=1$, again
similar to previous authors using this test case. The particles are
initially at rest and have an internal energy per unit mass of $u=0.05
\, G M / R$ and a ratio of specific heats of $\gamma=5/3$. 

Once released, the system undergoes rapid collapse from inside out and
reaches maximum compression at $t \approx 1.1$. Slightly before this
time, enough kinetic energy has been converted into heat to build a
pressure supported core in the center. Material that falls in later
bounces off this core and is accelerated outward, forming a strong
shock wave which interacts with the still infalling outer portion of
the sphere.

The simulations presented below were run using the Gadget MAC
(equation \ref{eq:gadget-mac}) for the tree with $\theta=10^{-4}$. We
used the spline kernel, equation \ref{eq:kernel}, with a fixed length
scale $h=0.02$ for softening the gravitational potential of the
particles. The hydrodynamic smoothing length was adaptive, so that
each particle retains $\approx 50$ neighboring particles (see section
\ref{sec:h_adapt}) at all times. The individual time step scheme
described in section \ref{sec:ind_ts} was used for all runs. The
parameters for the time step criteria (see equations \ref{force_crit},
\ref{vel_crit}, \ref{vel_acc_crit}, \ref{cfl} and \ref{eq:hchange})
were $\tau_{acc}=\tau_{vel}=\tau_{va}=1$, $\tau_{\mathrm{CFL}}=0.3$,
$\tau_h=0.15$. All runs were performed with the leapfrog integrator
and some of them with the Runge-Kutta-Fehlberg integrator as well,
with the latter using $\tau_{\mathrm{RKF}}=10^{-5}$ (see equation
\ref{eq:rkf-crit}). However, the difference between the two
integration schemes is very small (see below), so we focus on the
leapfrog simulations for most of the analysis presented here.

\subsubsection{Fixed Artificial Viscosity}

\begin{figure*}[!t]
   \epsscale{1.01}
    \plotone{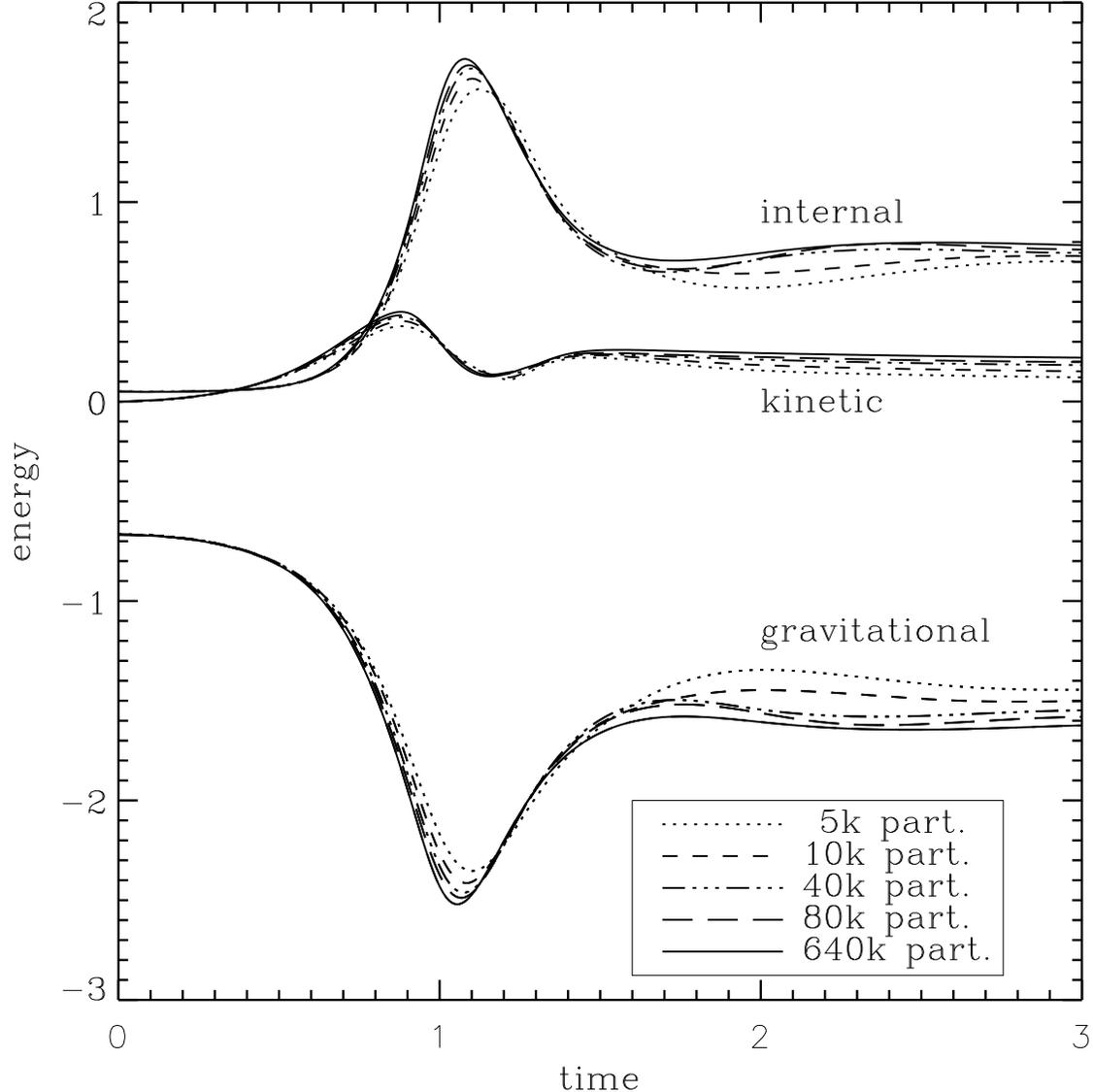}
    \caption{Time evolution of the total internal, kinetic and gravitational
    energy of the collapse test case. Resolution ranges from $5000$ particles
    to $80000$ particles, as shown in the box, with a run including $640000$ 
    particles run as a reference standard.}
     \label{fig_sph_energ}
\end{figure*}

\begin{figure}[!t]
   \epsscale{1.01}
    \plotone{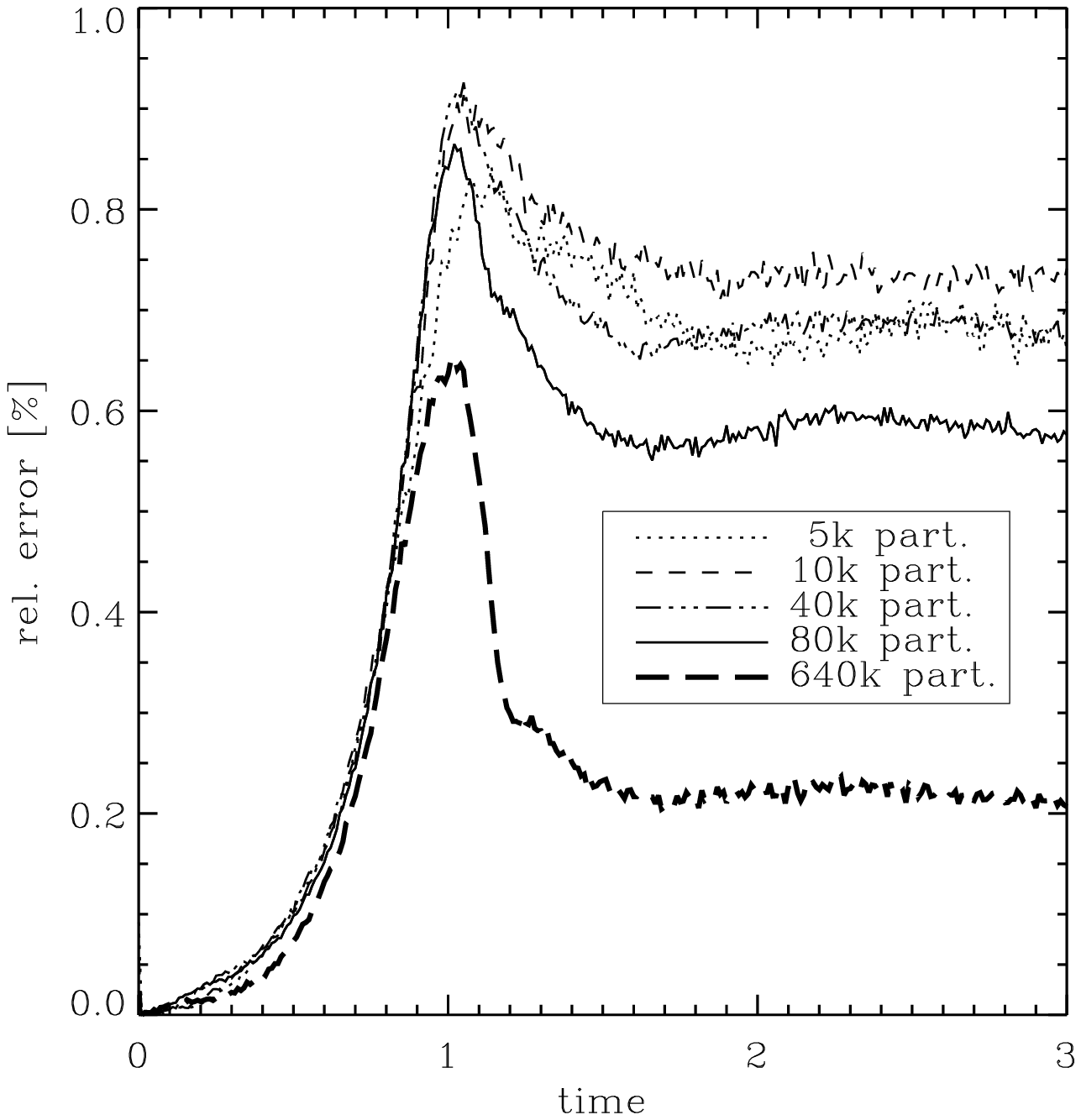}
    \caption{Time evolution of the relative error in total energy for
      the runs shown in figure \ref{fig_sph_energ}.}
     \label{fig_sph_energ_err}
\end{figure}

The evolution of potential, kinetic and internal energies of the system
is shown in figure \ref{fig_sph_energ}. The simulations have been
performed at four different resolutions, using $5000$, $10000$,
$40000$ and $80000$ particles. In addition, we have also run a
simulation with $640000$ particles as an internal reference standard
against which the other runs can be compared. This will make it easier
to assess the convergence of the simulations with increasing
resolution and the effect of resolution on dissipative effects.

Comparing the $40000$ and $80000$ particle runs around the time of
maximum compression ($0.7<t<1.5$), no energy component is different
from the corresponding value in the other simulation by more than a
few percent, with the largest differences coming near $t\sim1.1$, when
the gas is maximally compressed. Close examination of the curves show
that slightly larger differences between these two simulations and the
much higher resolution reference standard are present. For example,
near maximum compression, the reference run has converted somewhat
more gravitational energy into internal energy than any of the lower
resolution realizations. We believe this occurs because the
correspondingly lower dissipation at high resolution allows for higher
infall velocities and higher gas compression when the kinetic energy
is converted into thermal energy. For the same reasons, the peak
compression time occurs earlier as well. 

During the equilibration and late stage adiabatic expansion phase of
the system, after $t\approx 1.5$, differences between the $40000$ and
$80000$ particle runs relative to the $640000$ particle run become
more pronounced. At progressively higher resolution, more energy
remains in the form of internal energy, while the gravitational
potential energy remains more negative, indicating that the gaseous
core formed from the collapse is hotter and more tightly bound at
higher resolution than lower. The largest deviations between the
reference standard and either the 40000 or 80000 particle realizations
occur between $1.5 \le t \le 2$, and thereafter decrease. Even at
these times, the differences remain no larger than $\sim5$\%, and we
consider them to be well resolved, for the purposes required for this
test.

At the lower resolutions of $5000$ and $10000$ particles, the internal
and gravitational potential energies deviate from the $80000$ particle
run values by $\sim 10-15$\% at maximum compression, with even larger
systematic deviations at late times. These realizations have clearly
not reached the level of convergence to the proper solution that is
visible in the higher resolution runs.

The relative differences of a run integrated with the
Runge-Kutta-Fehlberg integrator (not shown) as compared to a
corresponding leapfrog run were never more than $0.63\%$ in the
internal energy, $0.44\%$ in the kinetic energy and $0.40\%$ in the
gravitational energy. The corresponding difference in total energy
reached a maximum of $0.32\%$. We do not consider these differences to
be highly significant and so in the remainder of our discussions, we
will present only the results of the simulations using the leapfrog
scheme. 

Figure \ref{fig_sph_energ_err} shows the relative error in total
energy as a function of time, for each of the four realizations at
differing resolution and our high resolution reference standard. In
every case, the errors peak near $t\approx 1.1$, corresponding to the
most compressed state of the system. Specifically, the maximum errors
in total energy are $0.84\%$, $0.91\%$, $0.93\%$ and $0.86\%$ for the
resolutions of $5000$, $10000$, $40000$ and $80000$ particles,
respectively. Although we have observed error peaks both in these
simulations and in those of other systems (see e.g. figure
\ref{fig_energ_v} below), their origin
remains somewhat unclear. One possibility is the fact that rather
large fluctuations in the neighbor counts of SPH particles over time
are permitted using the default settings of VINE. Similar fluctuations
have been shown to generate energy errors in other work \citep[e.g.
][]{AGW07} and may also contribute here. Rerunning our simulations
with tighter restrictions on the neighbor counts reduced the peak only
slightly however, so we could make no clear connection to this
possibility. 

Another potential source for the energy error is that the SPH
scheme in the current version of VINE does not incorporate the so
called $\mbox{\boldmath$\nabla$}h$ terms into the equations of motion
and energy. These terms arise from each particle having its own,
variable smoothing length and neglecting them can also give rise to
energy errors \citep{nel_pap93, nel_pap94}. However, these terms make
the SPH equations considerably more complex and have hence not been
widely adopted in the literature. A more promising and consistent
alternative is using the SPH formulation derived from variational
principles \citep{springel2002, monaghan2002}, which naturally takes
the $\mbox{\boldmath$\nabla$}h$ terms into account to all orders
without the terms being explicitly present in the equations
themselves. Such an approach will be included in a future release of
VINE. In the current revision, we note that we have been unable to
remove the error peaks completely through any combination of code,
integrator or force accuracy settings, though their magnitude changes
somewhat, while retaining computationally affordable simulations.
Nevertheless, the overall magnitude of the error peak is small
compared to ambiguities and shortcomings in the physical models of
most systems of interest in astrophysical contexts, and we believe
that the level of energy conservation produced by the code will be
sufficient to model such systems accurately.

After peak compression, the errors fall to $0.7\%$, $0.74\%$, $0.67\%$
and $0.57\%$, respectively, and remain nearly constant for the
remainder of the evolution. The $640000$ particle reference run
reaches much lower errors: $0.65\%$ at peak and $0.21\%$ during the
later phases of the simulation. Since both the integration accuracy
parameters $\tau_{i}$, $\tau_{\mathrm{CFL}}$, $\tau_{h}$ and
$\tau_{\mathrm{RKF}}$ and the MAC setting for the gravitational force
calculations were standard values and not particularly tuned for very
accurate time integration, we believe this level of energy
conservation is quite acceptable. Decreasing these parameters easily
leads to energy conservation to $\sim 0.25\%$ or better throughout the
simulation.

\subsubsection{Time dependent Artificial Viscosity}

\begin{figure*}[!t]
   \epsscale{1.05}
    \plotone{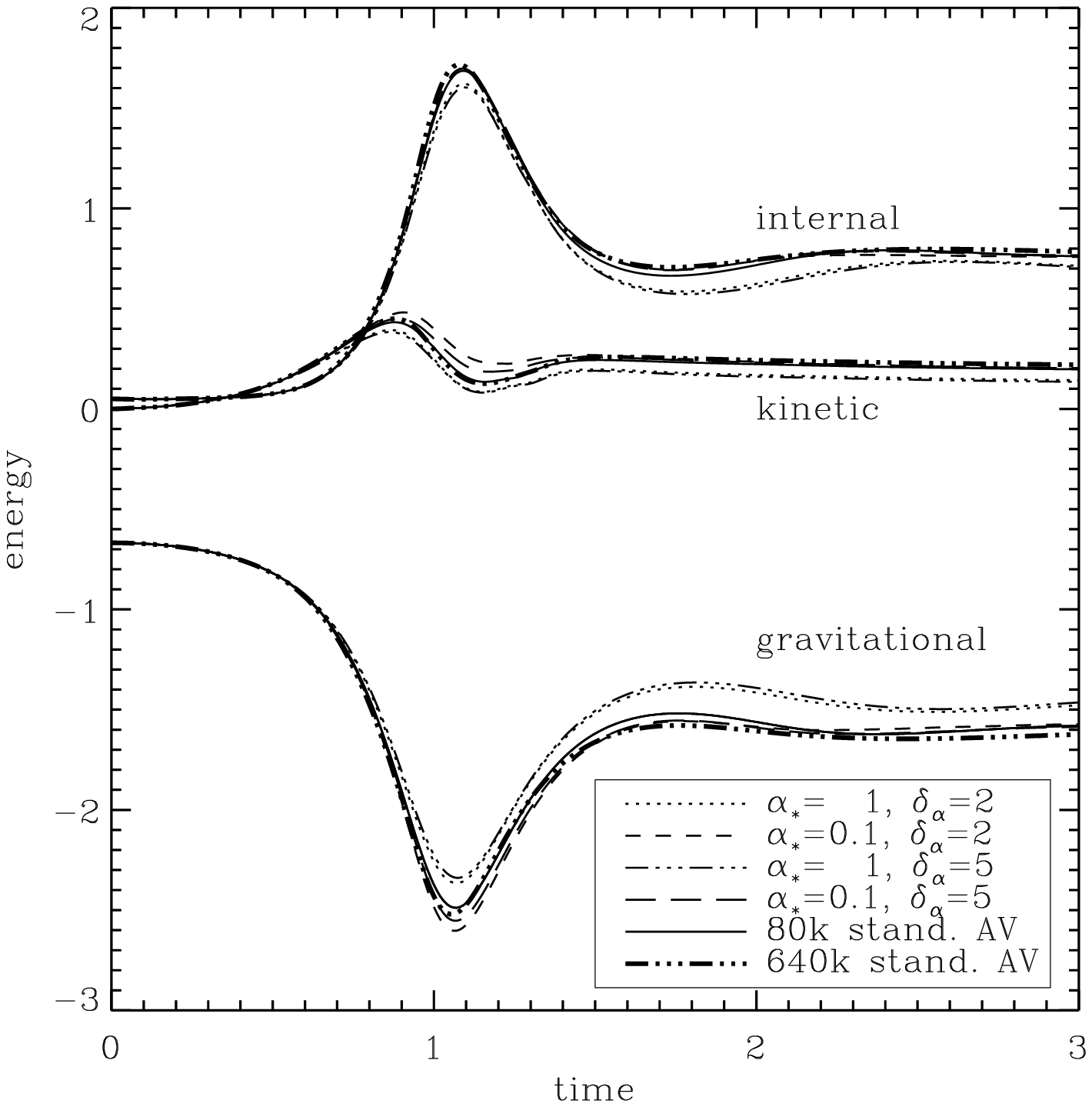}
    \caption{Time evolution of the total internal, kinetic and
    gravitational energy of the collapse test case for the model with
    $80000$ particles, using different implementations for the
    artificial viscosity. In addition, a high resolution reference run with $640000$
    particles is plotted as thick dash-dotted line.}
    \label{fig_sph_energ_av}
\end{figure*}

\begin{figure}[!t]
   \epsscale{1.05}
    \plotone{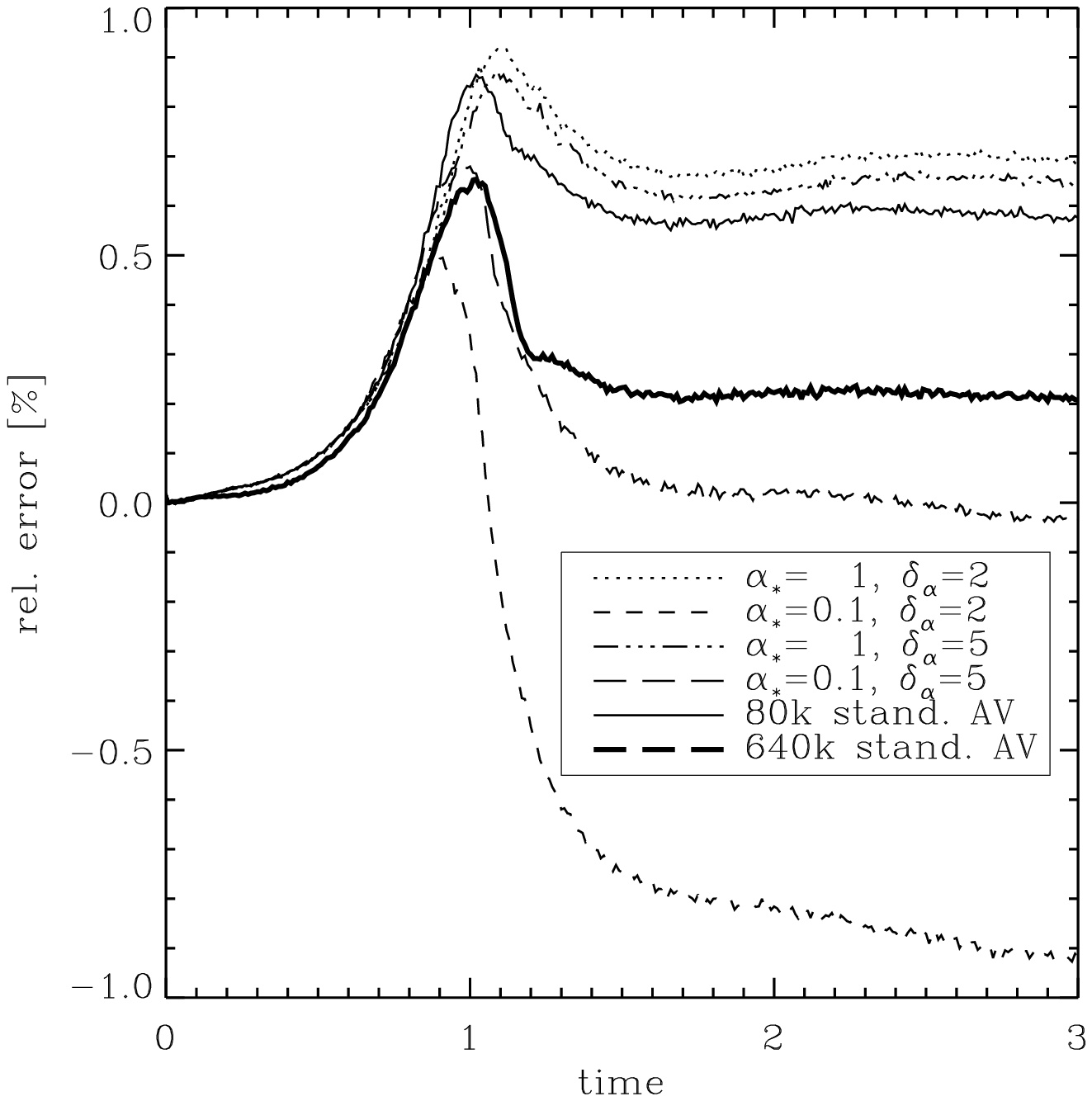}
    \caption{Time evolution of the relative error in total
    energy for the runs shown in figure \ref{fig_sph_energ_av}.}
    \label{fig_sph_energ_err_av}
\end{figure}

We have also used this test case to study the differences due to use
of the time dependent implementation of the artificial viscosity (AV),
as defined in equation \ref{eq:alpha}. In this test, we study
simulations at a single resolution of $80000$ particles and vary the
implementation of artificial viscosity. For reference and comparison
to the results above, we also include the same 80000 particle
simulation shown in figure \ref{fig_sph_energ}, characterized by an AV
formulation in its standard form (equation \ref{eq:avform}), with
coefficients fixed at $\alpha=1$ and $\beta=2$, and the Balsara
correction (equation \ref{form_func}). For comparison, we also include
the high resolution reference simulation, with the same formulation.
Models with the time dependent AV coefficients (equations
\ref{eq:alpha}--\ref{eq:s_term_av}) were run with different settings
of the parameters $\alpha_*$ and $\delta_\alpha$, defining
respectively, the value for the time dependent AV coefficient in
quiescent flow and the scaled length over which that coefficient
decays to its minimum value.

Figure \ref{fig_sph_energ_av} shows the time evolution of the total
internal, kinetic and gravitational energy of the system for the 80000
particle model without time dependent AV and for four models at the
same resolution identical except for changes in the time dependent
viscous coefficients. The distribution of energies among components
varies by as much as 10-20\% for different settings, with the two
$\alpha_*=1$ variations producing the largest differences during and
after peak compression. They are peculiar in the sense that
substantially less internal and kinetic energies are generated, than
in the fixed coefficient case. This is due to the higher dissipation
in the initial infall phase ($t<1$), which acts to slow the infall.
The behavior appears qualitatively similar to that seen in the lower
resolution variants in figure \ref{fig_sph_energ}, for which we expect
a correspondence between the lower resolution and higher dissipation.
Since, for the parameters used, the viscous coefficient will always be
larger than unity, the time dependent viscosity has the effect of
lowering the effective resolution in the simulation, but since it is
expected to be used only with a much smaller coefficient, the problem
will not be severe in practice.

Both models with $\alpha_*=0.1$ produce similar results for all three
energy components and, more importantly, results which are very
similar to the model with fixed AV. Only at and shortly after the
maximum compression do differences appear, most visibly in the kinetic
and gravitational energy components, where more resides in kinetic
energy than in the fixed AV case, and less (more negative) in the
gravitational potential energy. The differences from the $80000$ run
with fixed AV are at most a few percent over the entire duration of
the simulation. More importantly, the models with $\alpha_*=0.1$
resemble our high resolution reference run with $640000$ particles
better than the $80000$ particle run with fixed AV. The effect of the
time dependent AV, when used with values of $\alpha_*=0.1$, is to
simulate a higher resolution/lower viscosity system, and is clearly a
very desirable effect. Of the two models with $\alpha_*=0.1$, the one
with $\delta_\alpha=5$ is more similar to the $640000$ particle
reference run. At maximum compression, the minimum in the potential
energy is not as deep as in the model with $\delta_\alpha=2$, and the
correspondence between the kinetic energies is closer as well. At late
times, the model with $\delta_\alpha=2$ tends to higher (less
negative) potential energies, in fact higher than both the $640000$
particle run and also higher than the $80000$ run with fixed AV. 

In figure \ref{fig_sph_energ_err_av}, we show the relative error in
total energy for the four models with time dependent AV, the fixed AV
reference model and the high resolution reference run. The high
$\alpha_*$ models have larger errors than the model with fixed AV,
with the $\delta_\alpha=5$ model coming slightly closer to the fixed
AV result than the $\delta_\alpha=2$ model. The two low $\alpha_*$
models develop much smaller errors during maximum compression, but
after the peak, their energy errors go in the opposite direction. At
the end of the simulation, the error of the $\alpha_*=0.1$,
$\delta_\alpha=2$ model reaches $0.9\%$, as large or larger than the
peak errors of both the fixed high $\alpha_*$ models, and is still
growing. The model with $\alpha_*=0.1$, $\delta_\alpha=5$ shows the
best performance overall on this problem. Interestingly, its energy
error decreases after maximum compression, similar to the
corresponding $\delta_\alpha=2$ model, but becomes almost flat and is
$0.04\%$ at the end of the simulation. At and around the peak, its
behavior is very similar to the high resolution reference run. After
$t\approx 1.2$, it falls to lower errors than the high resolution run,
but overall it is still the model with comes closest to the $640000$
particle reference run. 

\begin{figure*}[!t]
   \epsscale{0.80}
    \plotone{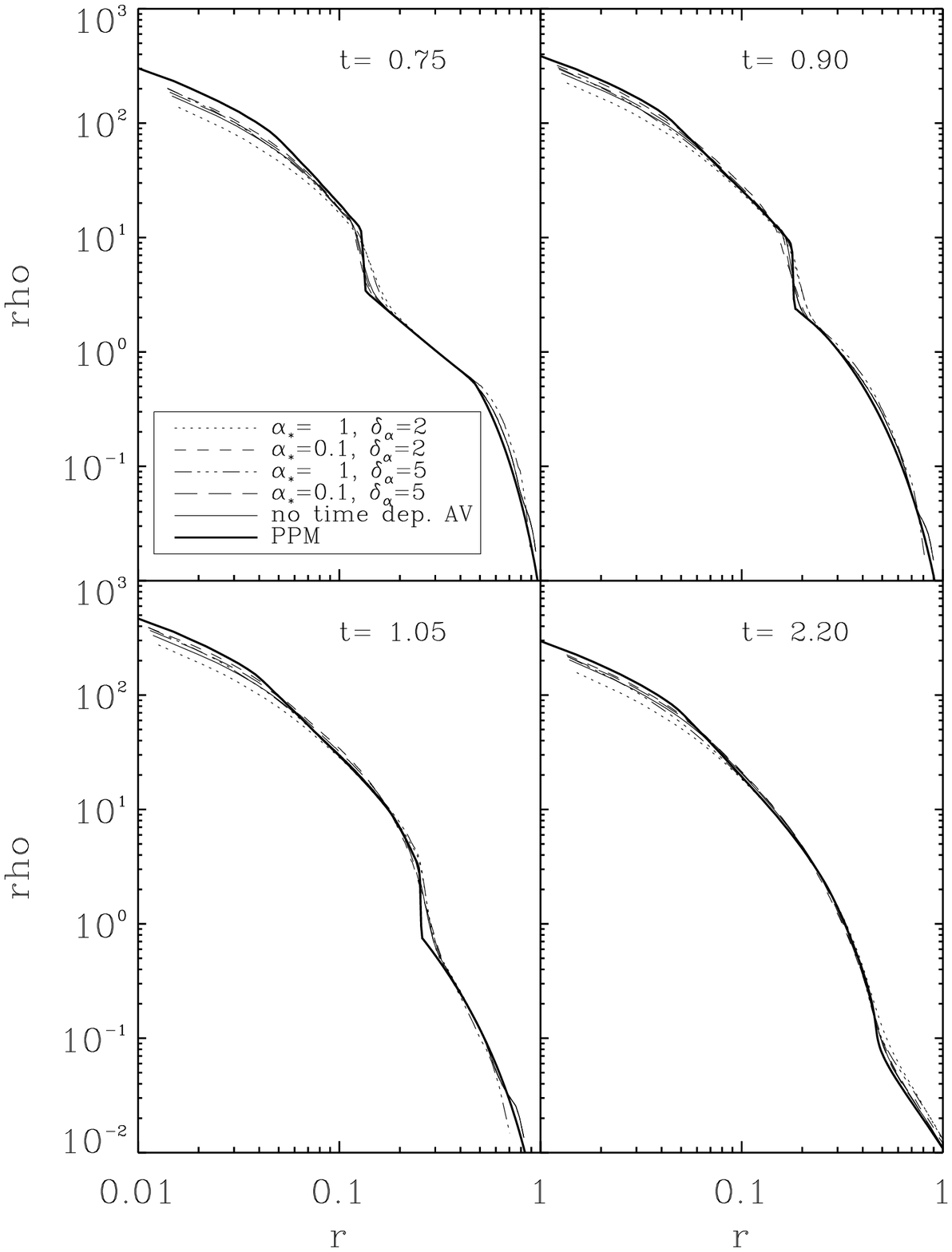}
    \caption{Radial density profile of the models shown in figure
    \ref{fig_sph_energ_av}. The thick solid line represents results of
    a PPM simulation by \citet{stein_muell93}. The upper two panels
    represent points in time before maximum compression in the center,
    the lower left shows the system at maximum compression and the
    lower right is after the shock wave has passed through most of the
    system.}
    \label{fig_sph_densprof_av}
\end{figure*}

\begin{figure}[!t]
   \epsscale{0.85}
    \plotone{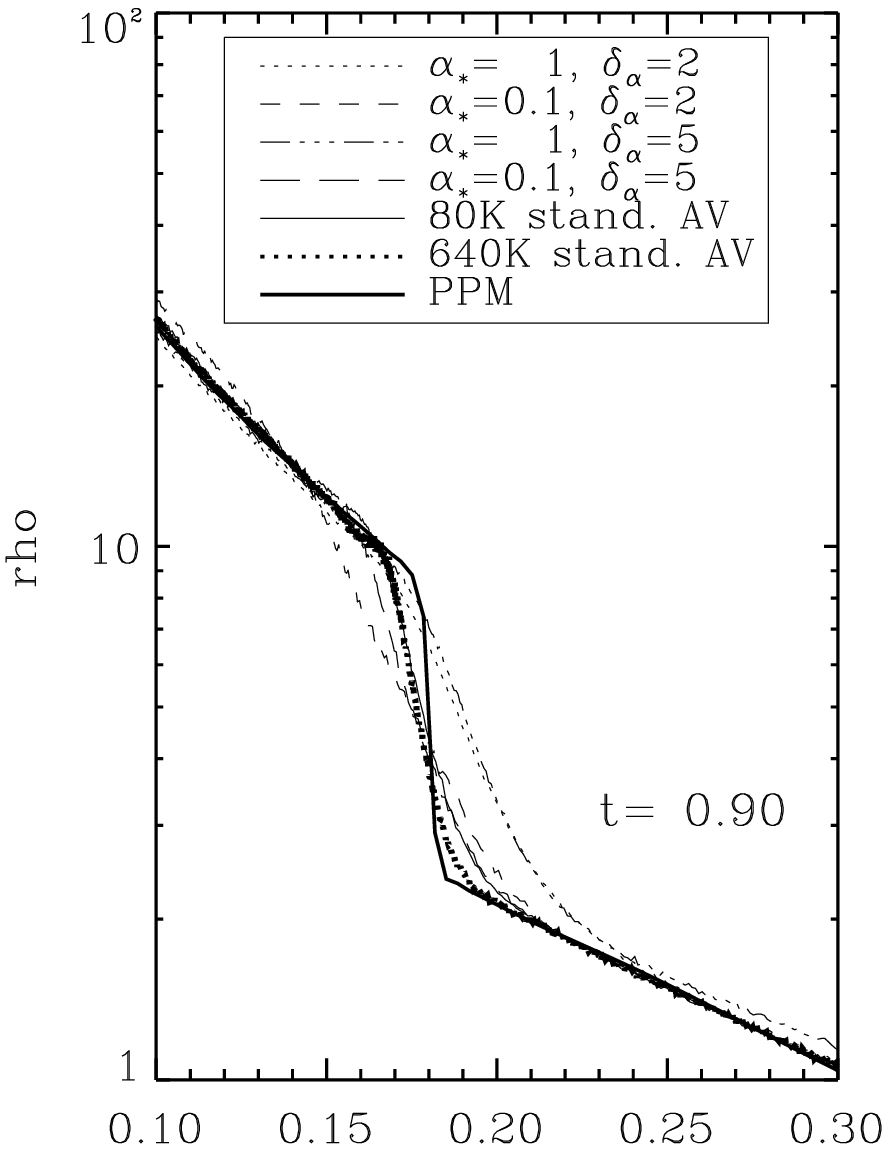}
    \caption{Radial density profile of the shock wave at $t=0.9$. The
    thick solid line represents results of a PPM  simulation by
    \citet{stein_muell93}.}
    \label{fig_sph_shock_av}
\end{figure}

\begin{figure}[!t]
   \epsscale{1.}
    \plotone{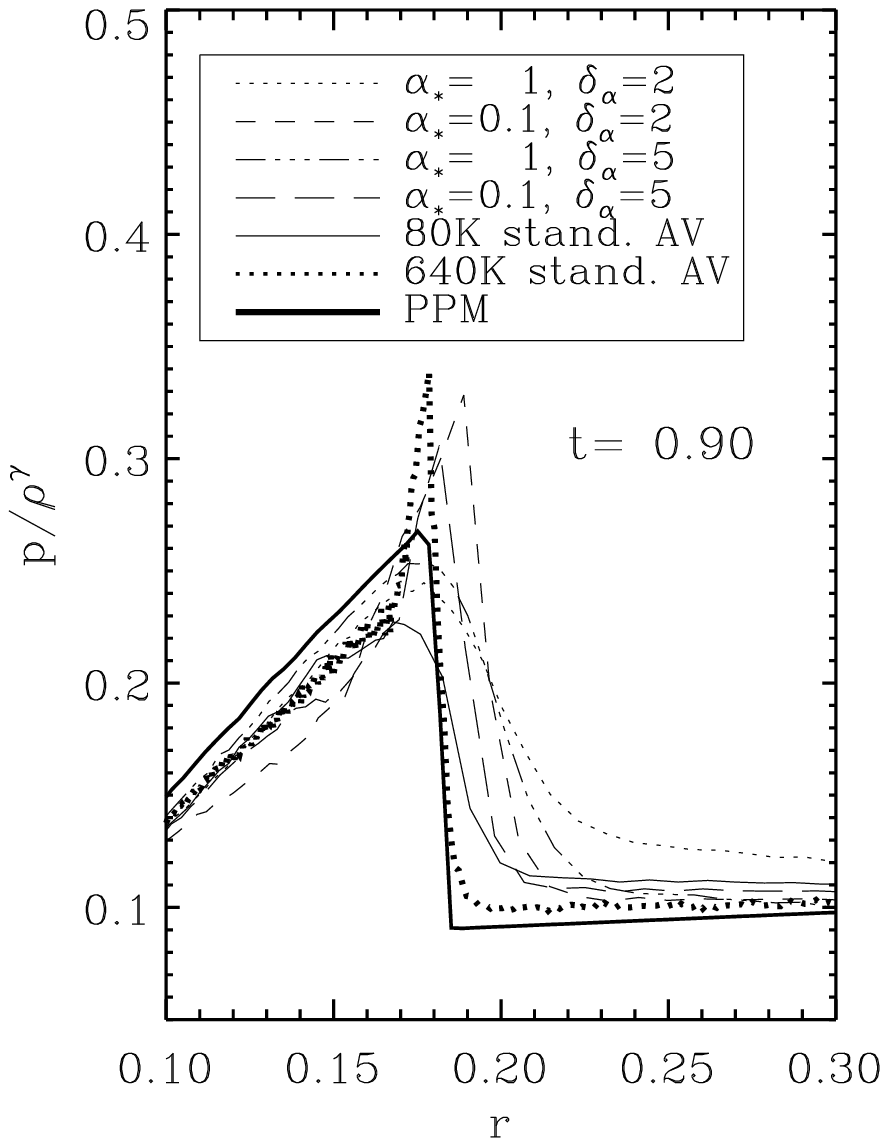}
    \caption{Radial entropy profile of the shock wave at $t=0.9$. The
    thick solid line represents results of a PPM  simulation by
    \citet{stein_muell93}.}
    \label{fig_sph_shock_av_entr}
\end{figure}

The differences seen in figures \ref{fig_sph_energ_av} and
\ref{fig_sph_energ_err_av} are due to the differences in the ability
of one set of AV parameters to model shocks and compressions with more
fidelity than another. Because modelling such shocks and compressions
is very sensitive to how AV is implemented in a code, we move now to a
closer examination of the radial density profiles of the simulations,
in which we expect a shock front to be clearly visible. 

Figure \ref{fig_sph_densprof_av} shows density profiles at four times during
the evolution for each of the models, and figure
\ref{fig_sph_shock_av} shows a close up view of the density at the
shock front at $t=0.9$. In figure \ref{fig_sph_shock_av_entr}, we show
the corresponding entropy profile in order to facilitate a comparison
with SPH variants which integrate an entropic function instead of
internal energy \citep[e.g.][]{springel_g2}. Pre-shock entropy
production is a known problem of SPH with this test case.

For comparison, the figures also show the result of a one dimensional
simulation of the same system using the Piecewise Parabolic Method
(PPM) with 350 zones, taken from \citet{stein_muell93}. We expect
better AV behavior in the SPH runs will be reflected in curves more
similar to that obtained from the PPM run. As an additional point of
reference, we also show the result of the $640000$ particle run in the
close up views of the shock front, shown in figures 
\ref{fig_sph_shock_av} and \ref{fig_sph_shock_av_entr}.

In every realization, the overall density profiles are similar, with
high densities in the center, a discontinuity further out, and a
decrease to low densities at the largest radii. The similarities give
confidence that the gross behavior of all variants produce sensible
results. There are differences specific to each realization however.
For example, the high viscosity models disagree most at the largest
radii, where the densities are overestimated relative to the PPM
model, especially at late times when the material is expanding freely.
The low viscosity and fixed AV models agree well with the reference
PPM curve there at all times. At small radii, all the SPH models
underestimate the central density. The low viscosity models
($\alpha_*=0.1$) agree best, followed by the fixed AV and long decay
constant, high viscosity model and, worst of all, the model with
$\alpha_*=1$ and $\delta_\alpha=2$, for which a peak density nearly a
factor two below that of the PPM run is observed. The central
densities agree with the PPM realization to within $\sim3.7\%$ in the
$640000$ particle reference run at $t=0.75$ (though are not plotted in
figure \ref{fig_sph_densprof_av}, in order to reduce clutter) and, at
late times, this SPH run actually overestimates the central density by
as much as 10\%. Using the central density as a metric, the models
with $80000$ particles used for this AV test are not fully converged,
but the similarity between the PPM and SPH methods at higher
resolution indicate that they do converge to very similar results,
given sufficient resolution. 

During the compression phase, the high viscosity models show
significant broadening of the shock front, as is expected, but also an
outwards position shift of the foot of the shock with respect to both
the PPM results and the other SPH models, especially visible in the
$t=0.9$ close up view in figures \ref{fig_sph_shock_av} and
\ref{fig_sph_shock_av_entr}. The low viscosity models reproduce the
density at the shock front more accurately and at the 
same radial position as the PPM model. In front of the shock front
(radially outwards), the densities in both low viscosity models are
very similar to each other and to the fixed AV model, while behind it, the
$\alpha_*=0.1$, $\delta_\alpha=2$ model exhibits a much shallower rise
than it should. The longer decay constant variant ($\alpha_*=0.1$,
$\delta_\alpha=5$) shows a much steeper density profile there, but still not
as steep as the fixed viscosity version, which appears to be the best
reproduction of the shock front. The high resolution model shows
closest correspondence to the PPM density curve at the foot of the
shock, but the high density side nearly over lies its lower
resolution cousin, also with fixed AV. Later, near maximum compression
($t=1.05$--lower left panel of figure \ref{fig_sph_densprof_av}),
simulations with all of the AV implementations misplace the shock
front, to slightly larger radii.

The entropy profiles shown in figure \ref{fig_sph_shock_av_entr}
exhibit behavior consistent with that shown for the density profiles.
The high viscosity settings (i.e. with $\alpha_*=1$) result in a wider
shock front than the low viscosity ($\alpha_*=0.1$) settings, which in
turn show the front at larger radii than the fixed AV settings. Of the
latter, the reference run with $640$k particles reproduces the PPM
reference location of the front very well. All of the time varying
viscosity settings generate a shock front at larger radii than either
the fixed viscosity settings or the PPM reference. All settings
exhibit the problem, common in SPH, of pre-shock entropy generation,
though at varying levels. The low viscosity models do slightly better
at radii $r\gtrsim0.22$, but worse at smaller radii, closer to the
front itself. Both also overpredict the entropy peak, similar to the
high resolution model with fixed AV. Behind the shock front, all
models produce lower entropies than the PPM reference run. The high
viscosity, longer decay settings are closest to the reference, while
the low/short settings underpredict by the largest margin. The
low/long model ($\alpha_*=0.1, \delta_\alpha=5$) most closely matches
both the low and high resolution fixed AV models interior to the
shock, recovers the shock position and width better than the settings
of any other time varying model and provides less pre-shock heating at
larger radii than the fixed AV model at the same resolution.

Given the behaviors we see in the shock structure and in both the
components of the energy and the conservation of total energy, we
conclude that the time dependent AV formulation allows a simulation to
be performed with dissipation from artificial viscosity at a level
comparable to that of a much higher resolution simulation that does
not include such a formulation. Our results show that the parameters
$\alpha_*=0.1$, $\delta_\alpha=5$ should be used for best results if
time varying viscosity is desired. Simulated with these parameters,
our test system yielded the best overall conservation of energy,
closest correspondence of all energy components to a high resolution,
fixed AV reference model, at the cost of a slightly more broadened
shock front compared to a run with fixed AV. Otherwise the shock
capturing abilities are comparable to the standard formulation with
fixed AV. In the absence of shocks, the time dependent AV formulation
yields better results than the time independent formulation. Therefore
we favor the use of the time dependent formulation.

These settings differ from those recommended by \citet{mor_mon97}, of
$\delta_\alpha\sim2$ for the decay length. Such differences illustrate
the problem dependent nature of the settings themselves, and are most
likely consequences of the presence of shocks of different strength in
different problems. Users of VINE who wish to employ the time
dependent AV formulation may need to account for such differences 
when choosing appropriate viscous parameters.

\subsection{Elliptical-Elliptical Merger Tests}\label{sec:drymerge}

The quality of hydrodynamic simulations with VINE has been studied in
detail in section \ref{sec:sphere-test}. In order to assess VINE's
ability to correctly evolve $N$-body models, we compare the
evolutionary trajectory of models simulated with VINE and with
Gadget-2. We use a merger simulation of two elliptical galaxies
\citep[a `dry merger', see e.g. ][]{naab2006} as a test problem. As
the gas fraction in such systems is very low, a pure $N$-body
representation of the galaxies, or one including only a small fraction
of gas particles, is a realistic model. The evolution of such a dry
merger requires correct behavior of the code over a wide range of
dynamical timescales. While the merger event itself is short compared
to the duration of disk galaxy mergers, its correct simulation
requires accurate time integration of particle trajectories in a
highly time dependent gravitational potential, requiring a wealth of
interesting dynamical behaviors to be accurately modeled. The correct
simulation of such a system is therefore a challenging test case.

The initial conditions for our tests are created by setting two
elliptical galaxies on a parabolic orbit, which is a reasonably
realistic setup for the orbit in a cosmological context
\citep{khochfar2006}. Each elliptical galaxy is understood to be the
remnant of a prior disk merger event whose components had a 1:1 mass
ratio and in which each of the spiral galaxies consisted of a stellar
disk and bulge as well as a dark matter halo, which in these tests
were modelled with $60000$, $20000$ and $120000$ 
particles, respectively. Therefore, each elliptical galaxy in our test
simulation consists of $160000$ stellar particles and $240000$ dark
matter particles so that the entire simulation contains $800000$
particles. For more details, we refer the reader to \citet{naab2006}.

\subsubsection{Requirements for sensible code to code
comparisons}\label{sec:sensible-settings}

For any physical model sufficiently complex to be interesting for
anything more substantial than an academic exercise, the only option
for validating the numerical code used to simulate it is to compare
against results obtained from other codes. In this sense, validation
effectively assumes that while those other codes have been developed
to solve the same set of equations, they have been constructed
differently enough, and have been validated independently based on a
sufficiently different set of criteria, to ensure that they are
effectively neutral standards against which the results of another
code can be measured. The question of whether or not this assumption
itself can be validated is unlikely ever to be answered
satisfactorily. Nevertheless, we will proceed as if it has been, and
compare the results of a simulation run using VINE with the results of
the simulation using the same initial conditions but evolved with the publicly
available and widely used Gadget-2 code of \citet{springel_g2}. 

Of course, various input settings used for a particular simulation will
directly affect both its accuracy and the overall results that are
obtained. It is trivial to run one or both codes with inappropriate
parameters for the given problem and thus find disagreement in the
results. Our first step in making comparisons must therefore be to
minimize the differences originating in different settings by running
an extensive set of tests prior to the actual comparisons that are our
main interest. Only when this process is complete can we attempt to
characterize results obtained from each code on an even handed basis.

For our comparisons, VINE uses its leapfrog integrator (section
\ref{sec_leap}) with individual timesteps (section \ref{sec:ind_ts})
and the Gadget MAC (equation \ref{eq:gadget-mac}). The Gadget-2 code
uses similar, but not identical, schemes for each of these three
choices. With these choices given, there still remains a very large
number of additional degrees of freedom defining all of the different
code settings available to the user, which are both different in each
code and, to the extent that a correspondence exists, may require
different values to produce results of similar accuracy. It serves
little purpose to match the settings of every one of these parameters,
especially if the normal settings for one or the other code are
significantly different. Instead, we choose to match the results of
the codes to what we believe are among the two most crucial
parameters: first the gravitational force accuracy, as controlled
through the MAC parameter $\theta$ (equation \ref{eq:gadget-mac}), and
second, the tolerance parameters, $\tau$, used in the integration
scheme (equations \ref{force_crit}--\ref{vel_acc_crit}). 

Even with only two parameters to adjust, some differences are
unavoidable. For gravity, VINE uses a multipole series truncated at
quadrupole order and the hexadecapole based Gadget MAC described in
\citet{springel2001}, while Gadget-2 uses a monopole truncated series
and a quadrupole based form of the MAC. Each code will therefore
require different values of the MAC parameter $\theta$ to obtain
similar force accuracies. When the mass distributions within a given
tree node are very inhomogeneous, as will certainly be the case in the
intermediate stage of this simulation, truncation errors in the
multipole summation will be systematically larger than for other
distributions, resulting in larger errors for a given setting. In
addition to the MAC differences, both codes test whether or not a
particle and node overlap in space, but use different procedures to
define approximate sizes for the nodes. While Gadget-2 errs
conservatively by requiring the particle to lie outside of a box which
is $\sim$20\% larger than the actual node \citep[see equation 19
in][]{springel_g2}, VINE uses a conservative definition of the node
size itself (see \vineII), which also errs by over-estimating the node
size. Which variant yields the more efficient criterion is not clear.

Looking at the time step criteria, Gadget-2 does not use timestep
criteria that include the velocity (equations
\ref{vel_crit}--\ref{vel_acc_crit}). Also the actual equation
implemented for the force criterion, equation \ref{force_crit},
included the tolerance parameter under the square root. So we expect
to find similar integration accuracy for both codes at different
values for the accuracy parameters. For simplicity in the discussion
below, we will denote the accuracy parameters with $\tau_{acc}$, but
actually include all three $\tau$ values in equations
\ref{force_crit}--\ref{vel_acc_crit} in VINE (set to the same value). We
will use the same symbol, $\tau_{acc}$, for Gadget-2 to mean the
single active accuracy $\tau$ parameter for Gadget-2, despite the
different functional form of its implementation.

\subsubsection{Determining comparable integrator and tree opening
parameters for each code}\label{sec:set-params}

We constrain the values of the parameters using a two step process.
First we determine node opening parameters which yield similar force
errors for the particles. For both codes, we require that $99\%$ of
the particles have force errors $\le 0.2\%$ and that the remaining
$1\%$ of the particles have gravitational force errors $\le 1\%$. To
some extent, these limits serve as proxies for the full error
distributions of all particle forces calculated by each code. More
importantly however, we note errors in the evolution of a simulation
can be dominated by large errors in only a few particles. In order to
eliminate such sources from consideration (or at least make them
similar in magnitude!), we require that each code produce similar
error limits (i.e. single points on the error distribution curves)
rather than similar error distributions. We use two limits in order to
ensure both that the full distribution of errors is sufficiently low
and also that a small tail of the distribution with high errors cannot
adversely affect the simulation outcome. 

For the second step, we use these MAC values in several full
simulations with each code, in which we vary $\tau_{acc}$ and we
require that the error in total energy never exceed $1\%$ at any time
during the simulation. While any particular choice of error limits is
somewhat arbitrary, some choice must be made. These values reflect
those used most frequently in our production runs. 

The setup was first simulated with VINE at very high accuracy. From
this reference run we selected snapshots at three different times
corresponding to the initial condition, just prior to the merger so
that one galaxy is located immediately adjacent to the other, and a
very late stage, well after the strongest interactions between the
galaxies are complete. We will refer to these three stages as the
initial, intermediate and late stages, respectively. For the VINE
calculations, accelerations for each snapshot were calculated directly
from the code, with the results saved to files for later analysis. In
order to compare identical mass distributions with the two codes, the
three snapshots of the reference run with VINE were converted to a
data format readable by Gadget-2. Each converted snapshot was read in
to the Gadget-2 code which calculated an independent reference
gravitational force for the particle distributions, as well as forces
at various values for the accuracy parameters, again saving the results for
later analysis.

For all three stages, we investigated the relative error in
gravitational forces as a function of the tolerance parameter $\theta$
used in the relevant MAC. Our procedure of comparing three different
stages of the simulation still does not necessarily guarantee that the
force accuracy over the entire simulation will remain comparable, but
we have some confidence that the deviations will not be too large
because we test three very different mass distributions,
representative of the range of mass distributions covered by the
simulation. 

\begin{figure*}[!t]
   \epsscale{0.95}
    \plotone{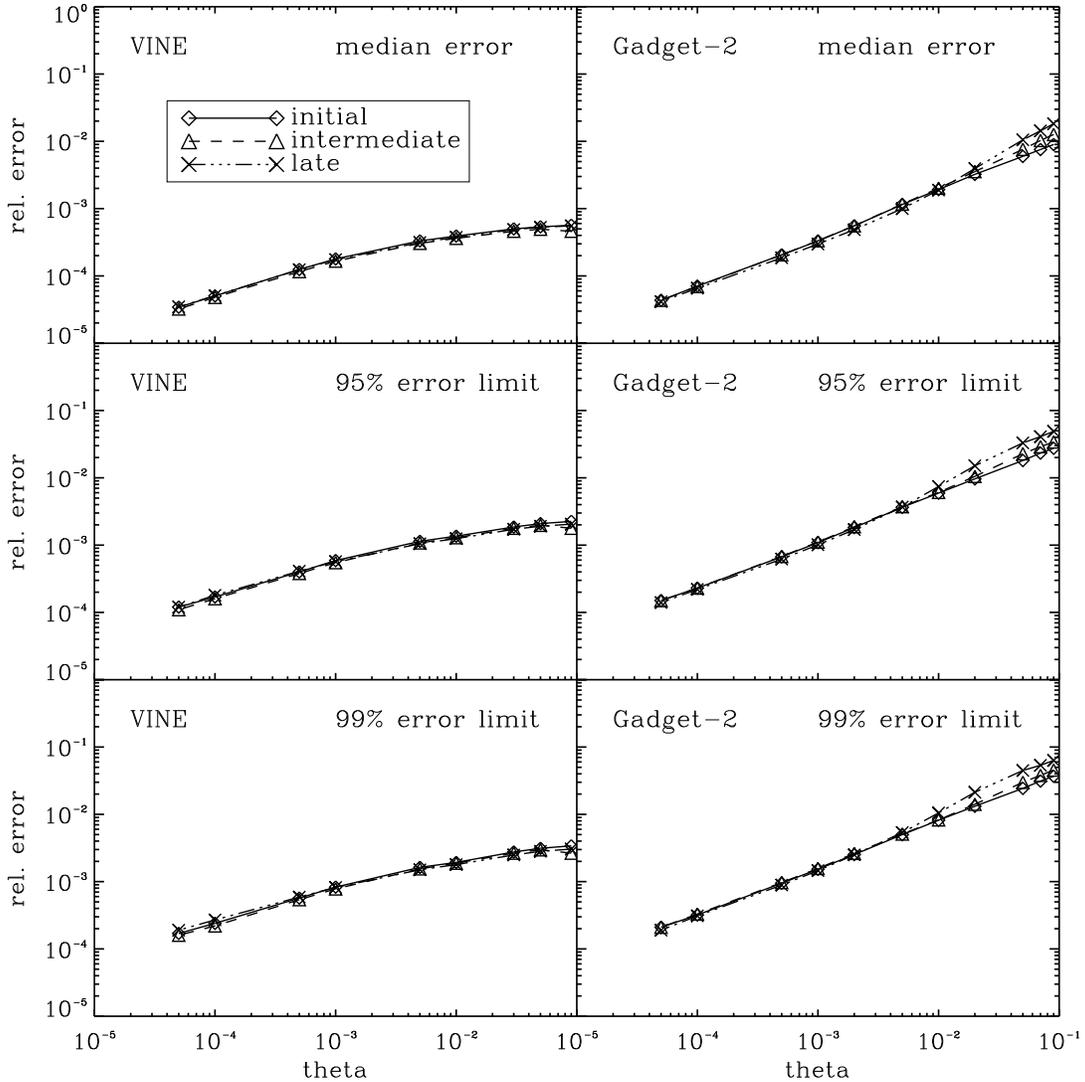}
    \caption{Relative errors of the gravitational forces as a function
    of tree node-opening criterion $\theta$ for VINE and Gadget-2. 
    From top to bottom, the median, 95\% and 99\% error values taken
    over all particles in each simulation. } \label{fig_err_theta}
\end{figure*}

Figure \ref{fig_err_theta} shows the error magnitudes for the median,
95\% and 99\% error limits for each code at three different times in
the simulations. As expected from the differences in the MAC
definitions, comparison of the error limits for the same value of the
opening criterion show that the limits for VINE are not the same as
those for Gadget-2, with the latter being considerably larger. The
differences are reflected in the required opening parameters: in order
to fulfill the error limits described above, we require a value of
$\theta=5 \times 10^{-3}$ for VINE and $\theta=10^{-3}$ for Gadget-2.
Note that the points corresponding to the next larger $\theta$ value
for which we determined force errors (i.e. $\theta=10^{-2}$ for VINE
and $\theta=5 \times 10^{-3}$ for Gadget-2) both still violate our
criterion of $99\%$ of the particles having force errors $\le 0.2\%$,
although the corresponding points in the bottom panels of figure
\ref{fig_err_theta} appear to be just below this limit. The parameter
value required for Gadget-2 lies well below that at which the late
time error curve diverges from the curves defining the initial and
intermediate time error limits. Its value is therefore not determined
by any peculiarities of the mass distribution which caused the
divergence in the first place and we do not expect that its value will
be significantly altered in other morphologies.

Beyond the simple statement of the MAC parameters required for each
code, it is of some interest to examine figure \ref{fig_err_theta}
further, to study the differences between the codes and, hopefully, to
better understand the consequences various choices of the criteria may
have on the simulations.

For VINE, the error magnitudes for all three limits follow similar
patterns as functions of the opening parameter, at all three times.
The error limit at the most permissive opening criterion for which we
obtained data terminates well below the 1\% level. Also, the error
limit curves appear to decrease their slope at larger $\log \theta$
values, rather than continuing a linear increase. Detailed tests
discussed in \vineII\  demonstrate that both phenomena continue
towards still larger $\theta$ values and that the error limits never
increase beyond $\sim 2-3$\%, even in the limit of a uniform density
(where we expect small force magnitudes due to the greater level of
cancellation of partial forces), a fact that will be very beneficial in
general for most simulations.

For Gadget-2, the error limit curves of the initial and intermediate time
particle distributions lie virtually on top of each other. The
late time curve deviates somewhat above $\theta > 10^{-2}$, exhibiting a
population of particles with force errors approaching
6\%--a level unacceptable for most simulations of astrophysical
interest and as much as a factor of 2 larger than either the initial
or intermediate time curves. Also, the error limits for a given
opening criterion do not appear to reach any limiting values as 
$\theta$ increases towards more permissive limits. These observations
are important because they demonstrate that the node opening criterion
must be chosen with some care in Gadget-2, if large errors are to be
avoided.

With the MAC parameters defined, we turn now to a set of simulations
in which we varied the integration accuracy $\tau_{acc}$, using the
same elliptical merger setup as above. Our criterion for sufficient
integration accuracy is to impose an upper limit on the error in total
energy. Adopting the force accuracy parameters $\theta$ from above and
using different values for the integration accuracy parameter
$\tau_{acc}$, we evolved our elliptical merger with VINE and Gadget-2
for 3.9 Gyr. However, as discussed in more detail below, we find that
the maximum energy error during the simulation with VINE depends not
only on the integration accuracy $\tau_{acc}$, but also on the force
accuracy parameter $\theta$. They span a two-dimensional parameter
space. So we adopt the $\theta$ values defined above as an upper limit
which guarantees to fulfill our force error criteria, but we explore
also the influence of lower $\theta$ values on the maximum energy
error occurring in the simulation for both codes.

As discussed in section
\ref{sec_ts_crit}, the choice of $\Delta t_{max}$ is effectively also
a time step criterion, so that $\Delta t_{max}$ and $\tau_{acc}$
actually span another multi-dimensional parameter space of computational
cost and accuracy. In this comparison however, we vary only the
$\tau_{acc}$ parameter, keeping $\Delta t_{max}$ fixed at $6.55$~Myr
for both codes, since we are interested in matching the simulation
parameters between the two codes to compare their results at similar
levels of error.

\begin{figure}[!t]
   \epsscale{1.0}
    \plotone{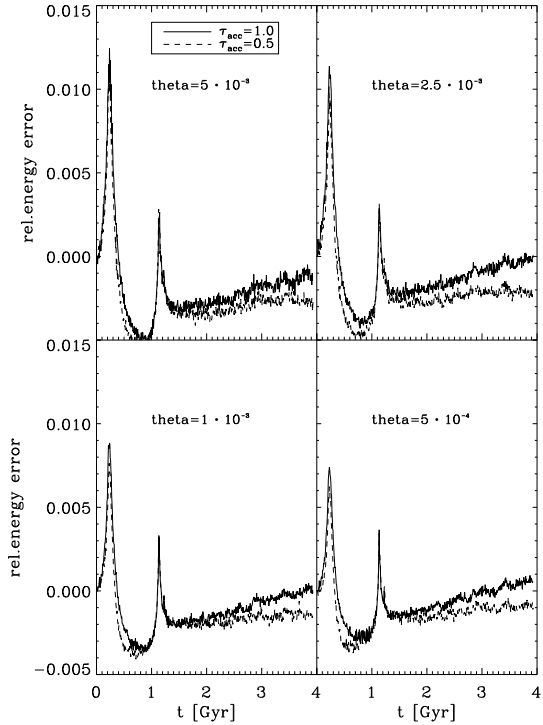}
    \caption{Relative error in total energy for an elliptical
    merger, using different integration and force accuracy
    parameters with VINE}
    \label{fig_energ_v}
\end{figure}

\begin{figure}[!t]
   \epsscale{1.}
    \plotone{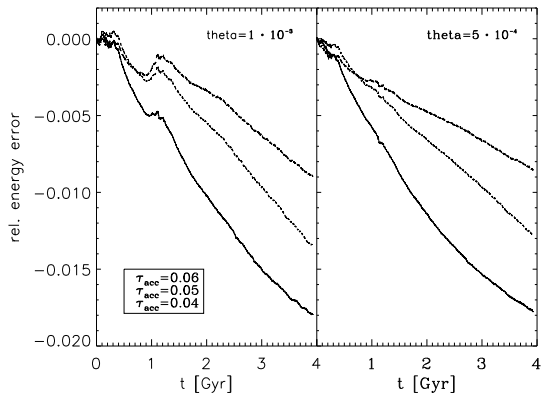}
    \caption{Relative error in total energy for an elliptical
    merger, using different integration and force accuracy
    parameters with Gadget-2.}  
    \label{fig_energ_g2}
\end{figure}

In figures \ref{fig_energ_v} and \ref{fig_energ_g2} we show the error
in total energy for $\tau_{acc}$ of 0.5 and 1.0 for VINE and 0.04,
0.05 and 0.06 for Gadget-2, respectively. The overall shape of the
energy error curves for the VINE simulations are quite different from
those for the Gadget-2 runs. While the VINE simulations show sharp
peaks at the times of the first close encounter and the final merger,
the simulations with Gadget lack a similar feature, although with
$\theta=10^{-3}$ a slight bump is visible. However, the Gadget
simulations exhibit a slow secular drift towards energy loss from the
system, with the magnitude of the loss being larger for the more
permissive integrator settings. 

The error peaks in the VINE simulations rise to $1.35\%$ for $\theta=5
\times 10^{-3}$ and $1.14\%$ for $\theta=2.5 \times 10^{-3}$. Only for
lower values of $\theta$ are the errors lower than our required limit
of $1\%$. If the integration accuracy parameter $\tau_{acc}$ is
lowered from 1 to 0.5, the peaks rise to slightly lower values, but
still only satisfy our limit of $1\%$ when $\theta$ is decreased to
$10^{-3}$. We have verified that still lower values for $\tau_{acc}$
than those shown in figure \ref{fig_energ_v} do not further decrease
the peak in the energy error. Hence the dominant effect which
determines the maximum energy error with VINE is the force accuracy
and not the integration accuracy. Obviously the two parameters span a
two-dimensional parameter space and in order to achieve low overall
error with VINE, decreasing the integration accuracy parameter
$\tau_{acc}$ only leads to better energy conservation if the force
accuracy parameter $\theta$ is set to low enough values. In order to
meet our criterion of less than $1\%$ error in total energy over the
course of the entire simulation, VINE requires a combination of
accuracy parameters $\theta= 10^{-3}$ and $\tau_{acc}=1$. 

For Gadget-2, there is also a dependence of the maximum energy error
on the force accuracy parameter $\theta$, but the dominant factor is
the integration accuracy parameter $\tau_{acc}$. Hence we keep the
force accuracy parameter $\theta=10^{-3}$ derived from the force
errors as above when investigating the integration accuracy. For
values of $\tau_{acc}=0.06$ and $0.05$, the maximum energy errors grow
beyond our limit of $1\%$. With $\tau_{acc}=0.04$ the maximum error is
$0.9\%$ and we adopt this value for our code comparison.

Finally, we also note that the secular energy drift in the Gadget
simulations may not be a problem of Gadget-2 itself, but rather that
the choice of $\Delta t_{max}$ is too large. In conjunction with the
differently implemented error criteria, this could permit a
sufficiently large number of particles to be integrated forward on
time steps too large to provide good fidelity. We have experimented
with smaller $\Delta t_{max}$ values for Gadget-2 and have found that
in fact, the drift can be alleviated with a smaller value, though at
the cost of increased computation. 

\subsubsection{Results of full scale merger simulations run with VINE
and Gadget-2}\label{sec:compare-result}

Using the above choices for the accuracy parameters, i.e.
$\theta=10^{-3}$ and $\tau_{acc}=1$ for VINE as well as
$\theta=10^{-3}$ and $\tau_{acc}=0.04$ for Gadget-2, we ran a full
scale elliptical merger model for $3.9$~Gyr with both VINE and
Gadget-2. The force errors and total energy errors for the two codes
were discussed in section \ref{sec:set-params}. Here we focus on a
comparison of the results of these simulations.

\begin{figure}[!t]
   \epsscale{1.}
    \plotone{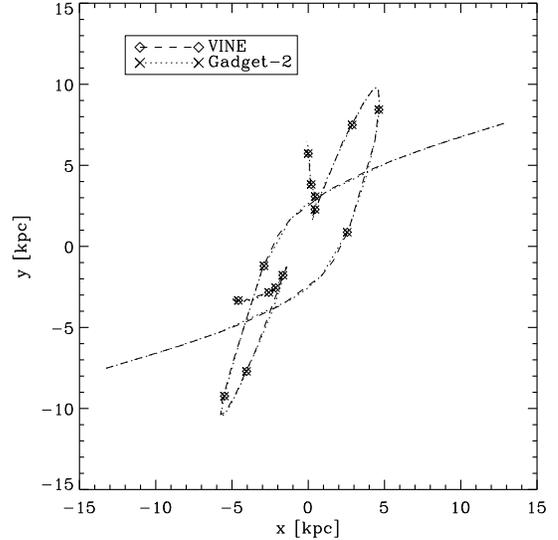}
    \caption{Center of mass trajectories in the orbital plane for the
      stellar component in each galaxy. The symbols along the
      trajectories represent the following times in Gyr: 0.25, 0.5,
      1.0, 1.25, 1.5, 2.0, 3.5}  
     \label{fig_cm_trajectories}
\end{figure}

A direct examination of the overall morphologies of the mergers at
different times, and as realized by the two codes, does not prove to
be of great use to compare differences in the evolution provided by
one code over the other. The features of the merger as evolved using
one are present in the other, and we are not able distinguish between
them in any quantitative fashion. Instead, and in order to compare the
results of both codes more quantitatively, we explore the behavior of
the centers of mass of the particles making up the stellar component
of each galaxy. The stellar component, as the most luminous, is most
interesting to compare because it resides in the central part of the
galaxy and dominates the potential there. Tracking its behavior
therefore means that we are tracking the dominant component of the
most dynamic region during the merger.

Figure \ref{fig_cm_trajectories} shows the center of mass trajectories
of the stellar component of both galaxies, as realized by each code.
The trajectories of the galaxies simulated with VINE and Gadget-2 lie
essentially on top of each other at all times. We have also plotted
symbols along the trajectories to indicate different epochs in the
evolution which have been chosen to sample the trajectories over a
wide range of evolutionary stages. The position of each snapshot along
the trajectories also matches between the Gadget-2 and the VINE
simulation.

In each case, the two galaxies have a first close encounter after
which they move to larger distances again. Then they reach a maximum
separation, turn around and approach each other again and finally
merge. During and after the final merger, the center of mass
trajectories of the stellar components do not converge to a single
point, even though the galaxies as a whole merge. Instead, they move
apart as a consequence of the asymmetric distribution and trajectories
of tidal debris well outside the newly formed elliptical galaxy (some
$\approx 15\%$ of the total stellar mass) which was created both
during the merger simulation and assumed in the progenitor
ellipticals, themselves taken to be remnants of mergers. The tidal
debris is accelerated outward during the merger event and thus the
center of mass of the star particles of a progenitor galaxy moves
gradually away from the center of the new elliptical, as the tidal
material moves to increasingly larger distances. The center of mass of
the system as a whole is of course not affected by this behavior.

\begin{figure}[!t]
   \epsscale{0.8}
    \plotone{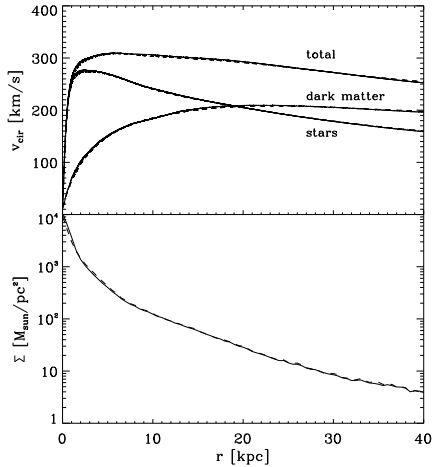}
    \caption{Rotation curve (upper panel) and surface density profile (lower
    panel) of an elliptical-elliptical merger. The solid lines are results
    using VINE for the simulation, the dashed lines show Gadget-2 results.}
     \label{fig_rot_ee_merger}
\end{figure}

We conclude our comparisons of the Gadget-2 and VINE simulations by
comparing the merger remnant produced by each simulation. In figure
\ref{fig_rot_ee_merger} we plot the rotation curve of the galaxy and
its surface density profile. The circular velocity profile of the
simulations lie nearly on top of each other for both stars and dark
matter, as do the surface density profiles over nearly their entire
range. Only in the outer regions of the galaxy, where resolution
begins to degrade, do very small differences become visible. Thus the
mass distributions of both the luminous and the dark component of the
merger remnant agree well between the two codes.

Given the complexity of the problem and the very different features
and algorithms implemented in the two codes, we conclude that both
codes agree very well on this demanding $N$-body test case.


\section{Performance of the Code}\label{sec:perf}

In \vineII \, we present detailed timings of the code on both serial
and parallel test cases. Various optimizations to the code are
described and their effects on the performance of the code
investigated. We will therefore leave our most detailed discussions of
the code's performance for \vineII, and perform only speed comparisons
between VINE and Gadget-2, using the same simulation used for the
accuracy comparisons in section \ref{sec:compare-result}. The
performance comparison to Gadget-2 will also serve as a frame of
reference for the timings presented in \vineII. 

\subsection{Speed comparison of VINE and Gadget-2}\label{sec:gadgcomp}

For this analysis we use snapshots of the simulations output by each
code at five different times during the simulation, to account for
variations in the speed due to the different mass distributions, seen
at different times. We adopt the same node opening criteria as in
section \ref{sec:set-params}, which yielded similar force accuracy for
each. Specifically, we used parameters of $\theta=1 \times 10^{-3}$
for VINE and Gadget-2, which guarantee
force errors of less than $0.2\%$ for $99\%$ of the particles. As
before, we convert an output dump from the VINE version of the
simulation to Gadget-2 format, and use that particle distribution in
the rate calculation in order to eliminate ambiguities due to
differences in particle distribution in the calculations of the rates.
For internal consistency, we also calculate separate `reference'
accelerations for each code using this particle distribution, for use
in the relative error determinations.

\begin{figure}[!t]
   \epsscale{1.}
    \plotone{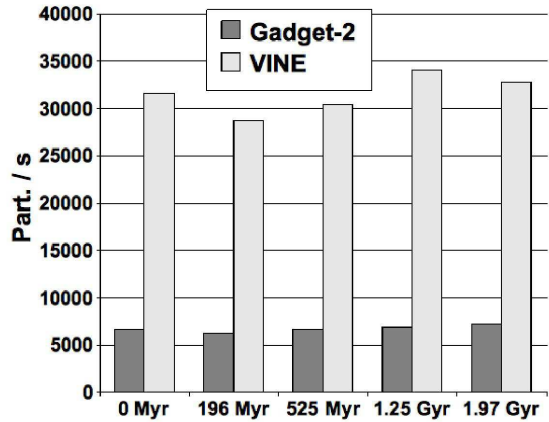}
    \caption{Calculation rate in particles per second for the
    gravitational force calculation of all $800000$ particles in an
    elliptical-elliptical merger simulation performed on one
    processor. The results are shown for different stages of the
    simulation.} \label{fig_gadg_vine1}
\end{figure}

\begin{figure}[!t]
   \epsscale{0.8}
    \plotone{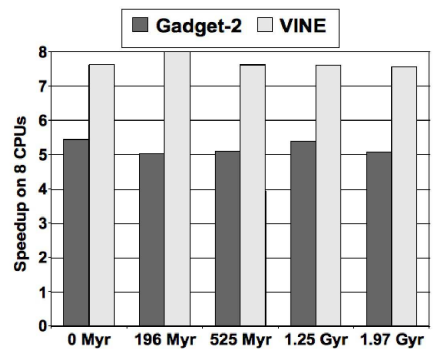}
    \caption{Speedup relative to one processor for the gravitational
    force calculation of all $800000$ particles in an
    elliptical-elliptical merger simulation performed on eight
    processors. The results are shown for different stages of the
    simulation. } \label{fig_gadg_vine8} 
\end{figure}

Figure \ref{fig_gadg_vine1} shows the speed of the gravitational force
calculation on all particles using these accuracy parameters on one
processor of the SGI Altix at the University~Observatory, Munich, i.e.
one Itanium~2 processor running at $1.5$~GHz. The snapshot at $t=0$ is
the initial setup, where the two galaxies are still well separated. At
$t=196$ Myr, the galaxies have their first close encounter. They pass
by each other and their distance increases until the turnaround point.
The snapshot at $t=525$ Myr is shortly before this point. Then the
two galaxies merge and form a new elliptical. The snapshot at $t=1.25$
Gyr is after the merger but before the system has had time to relax
completely. The final snapshot at $t=1.97$ Gyr is at a very late stage
when the newly formed system has already largely relaxed. For the five 
snapshots from $t=0$ to $t=1.97$ Gyr, the speed of the gravitational
force calculation using VINE ranged between $\sim 28600$ and $\sim
34000$ particles per second in serial model, while that using Gadget-2
between $\sim6200$ to $\sim7100$ particles per second. These rates
correspond to rate differences of a factor of $\sim$4.6--4.9 in serial
mode for VINE over Gadget-2.

The parallel speedup of the gravity calculation on eight processors,
relative to the serial calculation, is shown in figure
\ref{fig_gadg_vine8} for both codes at the same times as are shown in
figure \ref{fig_gadg_vine1}. In both cases, the scaling is very good.
At various times during the evolution VINE's speedup ranges between
$\sim7.6$ to just under a `perfect' scaling of a factor 8, while
Gadget-2's scaling ranges betwen $\sim5$ and $\sim5.4$. For the
eight processor calculation, these parallel speedups correspond to
rates ranging between $\sim 231200$ and $\sim 258900$ particles per
second for VINE, while Gadget-2 ranges over rates from $\sim 31300$ to
$\sim 37200$ particles per second. In terms of the difference in the
calculation speeds of VINE and Gadget-2, these speedups, yield a
difference of a factor of $\sim6.7-7.6$ for VINE over Gadget-2.

The scalings vary at different times in the simulation due to the fact
that the particle distributions themselves change over time, becoming
both clustered in the core regions of the galaxies and extended in
their tails. In turn, such clustering causes more work to be required
per update for some particles than for others or, in other words, it
causes load imbalance to develop between one processor and another as
this work is assigned unequally to each. For simulations of this
(comparatively small) size, such load imbalances will be more
difficult to overcome, since the total work is small. We see however,
that VINE's use of the OpenMP `dynamic' loop scheduling strategy to
assign small groups of particles to processors as they become idle,
provides somewhat better performance in this case than the code
specified distribution of particles provided by the distributed memory
strategy in Gadget-2. We expect that this difference will become
smaller for larger problem sizes and, perhaps, even be reversed at
very large sizes due to the synchronization overhead required for
OpenMP to parcel out additional work to many processors at once. Where
the crossover point occurs, and indeed whether it occurs at all, is
difficult to ascertain.

In addition to the timings of a single force computation at several
times, we have also measured the time for the full simulations. Both
codes were run in parallel on 8 processors of the same SGI Altix
machine used for the force calculation timings, on the same elliptical
merger simulation. VINE required 78445 seconds of wallclock time to
complete its version of the simulation, while Gadget-2 required 291076
seconds, a factor of $\sim3.7$ longer than VINE.

Both the snapshot and full simulation rates favor VINE over Gadget-2.
However, we recall that although we have attempted to use code
settings that yield similar error limits for both codes at all times,
differences between the resulting error limits will inevitably remain,
both between the two codes and for the same code at different times.
Therefore, before concluding that simulating merger evolution is
indeed faster with VINE than with Gadget-2, we must answer two
important questions. First, are the differences due to some
unanticipated bias in the code settings, differences in the error
limits obtained from each, or details of calculations themselves?
Second, why are the full simulation ratios smaller by a factor
$\sim1.9$, than the speed differences just quoted for a single gravity
calculation? 

A prime suspect in evaluating speed differences for full simulations
is whether or not the total number of force calculations performed
over the simulation is similar. Analysis shows that these numbers are
in fact very similar for both codes, $4.678 \times 10^9$ for VINE and
$4.391 \times 10^9$ for Gadget--a difference of only a factor
$\sim6$\% between the two. Therefore, the speed differences we see
cannot originate from this source. 

Next, we consider biases due to error limit differences. The node
opening parameter settings used in our test, figure
\ref{fig_err_theta} shows that the error limits are nearly identical
at all three times for both codes. Therefore we do not expect large
variation of the actual force errors during the simulation. However,
the gravitational error limits alone would allow for a slightly larger
value for the accuracy parameter $\theta$ in VINE and only the time
evolution of the error in total energy forced us to adopt a lower
value for $\theta$ for our timings.

An additional difference is that VINE always computes both the
gravitational potential and the force, while Gadget-2 computes only
the force. While it is possible for Gadget-2 to compute the potential
as well, it requires a second tree traversal and computation,
effectively doubling the time for the combined calculation. Such a
fundamental difference between the algorithms in the codes would
unfairly distort the comparisons between the timings in the two codes,
so we have compared the combined force and potential calculation of
VINE with the force only calculation of Gadget-2, except for snapshot
dumps, for which both are computed. We conclude that the calculation
of the gravitational forces in our test case is considerably faster
with VINE than with Gadget-2. 

Finally, we consider the differences between the speed ratios for the
full simulations and for the snapshot gravity calculations. The full
simulation timings include, of course, contributions from other code
operations than the gravity calculations themselves, such as the
leapfrog updates, and tree reconstruction and revision. They also time
the same overall activities in ways that are not captured by a single
snapshot timing, such as when only a fraction of particles are active.

Both VINE and Gadget-2 emit statistics on the costs of various
operations performed by the code, though the detailed metrics reported
in either case are not identical. According to the code-reported
timings for each, VINE and Gadget-2 require some $\sim35000$ and $\sim
255000$ seconds\footnote{For consistency with the single snapshot
calculations above, the latter value excludes a contribution from
calculations of the gravitational potential done by Gadget-2 for
snapshot dumps, in a separate step from the force calculations},
respectively, for `gravity only' portion of their total simulation
timings. These timings preserve the VINE/Gadget-2 speed ratio of $\sim
7$ reported for single time snapshots of the gravity calculations done
on 8 processors.

Full simulation timing differences must therefore originate in other
portions of the calculation. Specifically for VINE, the remaining
contributors to the total simulation time are dominated by the tree
rebuild and revision operations. Much of the tree revision cost can be
associated with systematic deficiency unique to the Altix
architecture. Specifically, we found in \vineII that parallel scaling
of the tree revision is particularly poor, saturating at a speedup of
only $\sim2$ even at very large processor counts. With scaling more
typical of that seen on other architectures VINE's performance would
no doubt improve.

Excluding the force calculation, the most costly activity in the
simulation is not tree revisions, but rebuilds. Of some relevance is
our conclusion in \vineII that the overall cost of tree rebuilds may
be amortized over more or less total simulation time by tuning the
frequency of rebuilds vs. revisions. The cost must be balanced against
an increase in the force calculation times however, and our timing
incorporates the settings, believed to be near optimal, recommended in
\vineII. To explore this question further, we reran the VINE
simulation with a more permissive setting (less frequent rebuilds) and
found that indeed while rebuild costs dropped substantially, force
calculation costs increased even more, resulting in poorer performance
overall.

We conclude with the observation that the merger simulation chosen for
our comparison study proves to be a remarkably challenging
configuration for both VINE and Gadget-2, but for different reasons.
For VINE, the tree rebuild and revision costs prove to be
substantially higher than we typically observe in simulations of other
physical systems, for which we typically find that they compose at
most 10-20
reduced parallel scalability in figure \ref{fig_gadg_vine8}, is
reported by the code itself to comprise approximately half of the
total run time. Similar scalability issues were originally discussed
by \citet{springel_g2} for a much smaller problem consisting of 60000
particles, and attributed to the fact that some particle
configurations prove to be quite challenging to the domain
decomposition algorithms used to distribute work across multiple
processors. 

Although the actual speed differences and scaling between the codes
will likely vary somewhat between different problems, we conclude that
VINE will be an excellent choice for use in efficiently solving most
problems encountered in astrophysical contexts.


\section{SUMMARY}\label{sec_summary}

In this paper we have introduced VINE, a hybrid $N$-body/SPH code
which uses a binary tree structure for the calculation of
gravitational forces. It is a very modular and flexible code which
allows the user to compile and use only those modules which are
required for simulating the physics of the problem at hand. This
modular structure also makes it fairly easy to exchange such modules
for new ones if needed or to add others to implement new physics or
numerical features. 

The code includes both a Runge-Kutta-Fehlberg and a leapfrog
integrator, which can be chosen by the user at compile time. Both can
make use of an individual particle timestep scheme. We have described
the SPH implementation in VINE, including details of the
symmetrization and the scheme for adapting the smoothing lengths. VINE
includes the capability for calculating gravitational or other long
range forces using a tree structure to organize and sort
particles into near and far interactions, and an outline of the
techniques are described here. We describe the implementation of
periodic boundary conditions used in both the gravity and the SPH part
of the code. Finally, we demonstrate the capability of the code to
accurately simulate both hydrodynamic and $N$-body problems using, in
the first case, the collapse of a gas sphere as a test problem and, in
the second, an elliptical-elliptical galaxy merger. 

We have demonstrated that the code performs well on a standard, but
somewhat contrived, test problem with a well known result--the Evrard
collapse problem. More importantly, we demonstrate that it performs
well on a full astrophysical simulation of the merger of two
elliptical galaxies, in comparison to the publically available
Gadget-2 code. The performance of the gravitational force calculation
in this test, the most costly component of simulations including
self-gravity, was superior to that of Gadget-2. The speed difference
ranged between $4.6$-$4.9$ times faster on one processor at different
times in the calculation, with better parallel scaling than provided
by Gadget-2 as well. For a full simulation timing on 8 processors, we
find a smaller difference, with VINE being only $\sim 2.9$ times
faster.

\subsection{Additional optimizations, features and future directions}

VINE, as it has been presented here and in \vineII, will be released
to the public under GNU Public License. We hope that it will become a
useful tool for use on a wide variety of problems. At present, the
code exists in a flexible, `base' version, which includes a number of
basic physics packages common in many astrophysical contexts, but is
by no means complete. We expect that other workers will wish to
incorporate packages not included in the base version, or to make
changes to it in order to advance their own research goals. As one of
its strengths for users, we expect that this process will not be
overly burdensome to its users, due both to the features included in
its base version and the overall modular design. In the future, we
expect to push the code to become still more of a more general purpose
`multi-physics' code, for use in a wider variety of contexts. 

Already, efforts are underway to incorporate such additional physics
as required to satisfy our own research goals, for example. At the
University Observatory in Munich, efforts are underway to develop
modules to implement radiative cooling, subgrid models for star
formation on galactic scales, stellar feedback, black hole accretion,
AGN feedback, inflow boundaries and ionizing radiation. These packages
are at many different stages of development, with some essentially
complete, others are merely in planning stages, while still others,
such as a module to simulate cosmological evolution, exist in the code
already but are not well maintained and may no longer be fully
functional. VINE has also been the basis for an implementation of the
SPH method on special purpose, reconfigurable hardware, so called FPGA
(\ti{F}ield \ti{P}rogrammable \ti{G}ate \ti{A}rray) boards 
\citep[see e.g.][]{fpa2007}.

In current form, VINE's SPH module does not include strength and
damage models needed to simulation systems including solid bodies. In
future work, we intend to integrate into VINE such a model, present in
a much earlier cousin \citep{BA95}, for use in modern simulations of,
e.g., giant impacts between planetary bodies or of asteroids on earth.
These types of calculations may also require different or more
advanced treatments of the hydrodynamics, such as are available using
the Moving Least Squares Particle Hydrodynamics (MLSPH) approach of
\citet{Dilts99,Dilts00}, which may be substituted for VINE's standard
SPH module. 

Other features of interest include alternate integrators and
improvements to the physics modules that already exist in the code.
The VINE framework could, for example, be adapted to accommodate
integrators specially designed for use in highly accurate $N$-body
simulations \citep{aarseth99,levdunc94}. A KDK variant of the leapfrog
integrator \citep{springel_g2} is also of great interest, due to its
stability properties and the fact that force calculations can be
parallelized with higher efficiency when used in individual time step
mode. Similarly, modifying the SPH module to incorporate an entropy
conserving formalism similar to that of \citet{pm04} will permit a
more faithful reproduction of hydrodynamic features of the flow.

VINE has been best tested on problems characterized by moderate
particle counts (up to a few $\times 10^6-10^7$) but very large time
step counts per particle during the simulation lifetime, by some
fraction of the total number of particles. Common examples of such
problems are simulations of systems with widely varying dynamical time
scales, such as occur in the evolution of circumstellar disks or in
galaxy collisions. Because the large time step counts occur for only
the comparatively small fraction of particles defining the most
dynamically active regions, only a small fraction of the particles
require new force calculations at a given time and the code's parallel
efficiency suffers. An important consequence of this effect is that it
frequently precludes simulations with very large particle counts, due
to the accompanying increase in time to completion, and is a major
motivation behind our effort to optimize VINE for speed and fine
grained parallelism, rather than to explore its performance at more
extreme computational scales.

Indeed, at the largest scales where computing platforms largely
consist of clusters of nodes each with their own private memory, VINE
cannot be run at all in its current form. It will therefore be a
priority for future versions of VINE to include a distributed memory
layer, most likely using standard `Message Passing Interface' (MPI)
library functionality, in order to be more generally useful in cluster
environments as well. The current version of VINE is best optimized
and tested for moderate sized, shared memory machines with up to
$\sim100$ processors. Currently, multi-core processors and multi-chip
nodes are becoming common in inexpensive, commodity computing
environments. VINE will be well suited to these machines on problems
up to at least tens of millions of particles. Larger scale shared
memory machines are available in sizes up to hundreds to thousands of
cores, and we expect that VINE will run unaltered on all or parts of
such machines, though users may encounter issues on such scales that
we have not explored.

We expect that if VINE finds wide use in the astrophysical community,
many other modules may be developed, beyond those suggested here. We
hope that the code will be as useful to others in the astrophysical
community as it has been for us so far.

\subsection{Availability of the code}\label{sec:wrapup}

The code is available to the public under GNU General Public License
version 2, via download from the ApJS website, directly from the
authors or via download at the USM website:
http://www.usm.lmu.de/people/mwetz and at both
http://arXiv.org/abs/0802.4245 and http://arXiv.org/abs/0802.4253, by
choosing the "Other Formats" sublink and then "Download Source".


\acknowledgements
We wish to thank Willy Benz for his generous gift to so many, over so
many years, of his SPH wisdom and the original code on which VINE is
based. We thank the referee, Volker Springel, for many useful
suggestions which improved the quality of the paper. Some of the
computations reported here were performed on the 
SGI Altix 3700 Bx2 supercomputer at the University Observatory, 
Munich, which was partly funded and is supported by the DFG
cluster of excellence "Origin and Structure of the Universe"
(www.universe-cluster.de). Other computations and most of the code
development used facilities at the UK Astrophysical Fluids Facility
(UKAFF).  Portions of this work were carried out under the auspices of
the National Nuclear Security Administration of the U.S. Department of
Energy at Los Alamos National Laboratory under Contract No.
DE-AC52-06NA25396, for which this is publication LA-UR 08-0429.  We
thank S. Khochfar and M. Bertschik for their initial work on the
implementation of the periodic boundaries. We thank Volker Springel
for help with the Gadget-2 code as well as making a fix for a problem
with the domain decomposition available to us. We thank Matthias
Steinmetz for making the PPM results of the collapsing sphere test
available to us. MW acknowledges support by Volks\-wagen Foundation
under grant I/80~040. AFN wishes to thank UKAFF for financial support. 

\bibliographystyle{apj}
\bibliography{lit}

\end{document}